\documentclass[aps,prd,showpacs,notitlepage,nofootinbib,preprintnumbers,amsmath,amssymb]{revtex4-1}

\usepackage{graphics,graphicx}
\usepackage{dcolumn}
\usepackage{bm}
\usepackage{epsfig}
\usepackage[usenames]{color}
\usepackage{hyperref} 
\usepackage{mathbbol}
\usepackage{epstopdf}
\usepackage{simplewick}
\usepackage[utf8]{inputenc} 
\usepackage{slashed}
\usepackage{stackrel}

\def\eq#1{{Eq.~(\ref{#1})}}
\def\fig#1{{Fig.~\ref{#1}}}
\newcommand{\ben}{\begin{eqnarray*}}
\newcommand{\een}{\end{eqnarray*}}
\newcommand{\un}[1]{\underline{#1}}

\newcommand{\pd}{\partial}

\newcommand{\tr}{\mbox{tr}}
\newcommand{\thalf}{\tfrac{1}{2}}

\newcommand{\as}{\alpha_s}

\newcommand{\dhd}{{\textstyle d}
\lower.03ex\hbox{\kern-0.38em$^{\scriptstyle-}$}\kern-0.05em{}}
\newcommand{\dbar}{{\textstyle \delta}
\lower.03ex\hbox{\kern-0.38em$^{\scriptstyle-}$}\kern-0.05em{}}
\newcommand{\half}{{1\over 2}}

\newcommand{\bra}[1]{\left\langle #1 \right|}
\newcommand{\ket}[1]{\left| #1 \right\rangle}

\newcommand{\ul}[1]{\underline{#1}}

\begin{document}

\title{Valence Quark Transversity at Small $x$}

\author{Yuri V. Kovchegov} 
         \email[Email: ]{kovchegov.1@osu.edu}
         \affiliation{Department of Physics, The Ohio State
           University, Columbus, OH 43210, USA}

\author{Matthew D. Sievert}
  \email[Email: ]{sievertmd@lanl.gov}
	\affiliation{Theoretical Division, Los Alamos National Laboratory, Los Alamos, NM 87545, USA}

\begin{abstract}
In our previous work we established a formalism which allows one to determine the small-$x$ asymptotics of any transverse momentum-dependent parton distribution function (TMD PDF) of the proton at small values of strong coupling. In this paper we apply this formalism to the valence quark transversity TMD. We relate the valence quark transversity to the transversely polarized dipole scattering amplitude, written in terms of the fundamental transversely-polarized ``Wilson line" operator, an expression for which we derive explicitly as well. We then write down the evolution equation for the transversely polarized dipole amplitude. Solving that equation we arrive at the following small-$x$ asymptotics of the valence quark transversity in the large-$N_c$ limit:
\begin{align}
h_{1T}^v (x, k_T^2) \sim h_{1T}^{\perp \, v} (x, k_T^2) \sim \left( \frac{1}{x} \right)^{-1 + 2 \, \sqrt{\frac{\alpha_s \, N_c}{2 \, \pi}} } . \notag
\end{align}
This result is in agreement with one of the two possible small-$x$ asymptotics for the transverse structure function found previously by Kirschner, Mankiewicz, Sch\"{a}fer, and Szymanowski in \cite{Kirschner:1996jj}.
\end{abstract}


\maketitle

\tableofcontents



\section{Introduction}

Quark transversity TMD is an interesting and important object to study \cite{Ralston:1979ys,Cortes:1991ja,Jaffe:1991ra,Barone:2001sp,Anselmino:2007fs}. It provides a key insight into the distribution of spin among the quarks in a transversely polarized proton. Moreover, the $k_T$-integral of the quark transversity TMD is equal to the chiral-odd transversity PDF $h_1 (x, Q^2)$. In turn, the transversity PDF $h_1$ is related to the tensor charge of the proton \cite{Jaffe:1991ra},
\begin{align}\label{tensor}
\delta q (Q^2) = \int\limits_0^1 dx \ \left[ h^q_1 (x, Q^2) - h^{\bar q}_1 (x, Q^2) \right],
\end{align}
a fundamental quantity in quantum chromodynamics (QCD) which is also employed in the searches of physics beyond the Standard Model. Recent phenomenological efforts to extract the proton tensor charge from a global fit to the available world data \cite{Radici:2018iag,Bacchetta:2012ty} led to a significant tension between the obtained tensor charge for up ($\delta u$) and down ($\delta d$) quarks and the results of lattice simulations. A possible resolution of this ``transverse spin puzzle" is the possibility for small-$x$ quarks to carry a significant amount of transverse spin. If the quarks at values of $x$ smaller than those measured in experiment carry a large amount of transverse spin, then the discrepancy between the tensor charge values extracted from experimental data and the lattice results may be accounted for. In this paper we will try to understand the plausibility of this scenario by theoretically determining the small-$x$ asymptotics of the quark transversity distribution. 

In our previous paper \cite{Kovchegov:2018znm} we developed a general method for determining the small-$x$ asymptotics of quark TMDs. The developed formalism was applied to quark helicity TMD, reproducing the results of the earlier works on the subject \cite{Kovchegov:2015pbl,Kovchegov:2016weo,Kovchegov:2016zex,Kovchegov:2017jxc}. (Gluon helicity was considered separately in \cite{Kovchegov:2017lsr}.) The technique developed in \cite{Kovchegov:2018znm} consists of the following steps: (i) starting with the operator definition of the TMD, simplify it for the case of small $x$, rewriting the TMD in terms of the ``polarized dipole operator", involving the so-called polarized Wilson lines --- operators made of semi-infinite and finite light-cone Wilson lines with one or more sub-eikonal operator insertions; (ii) simultaneously with the first step, or after it, one has to construct the explicit expressions for the relevant polarized Wilson line operators; (iii) construct small-$x$ evolution equation (or equations) for the polarized dipole amplitude. These equations usually close in the large-$N_c$ and/or the large-$N_c \& N_f$ limits. (iv) Solving the evolution equations one obtains the small-$x$ asymptotics for the polarized dipole amplitude, and, through the relation found in step (i), for the quark TMD in question. In the case of helicity evolution, this method has been applied to the leading small-$x$ contribution \cite{Kovchegov:2015pbl,Kovchegov:2016weo,Kovchegov:2016zex,Kovchegov:2017jxc}, which results from resumming powers of $\as \, \ln^2 (1/x)$. This resummation parameter was originally suggested in \cite{Kirschner:1983di} for certain types of small-$x$ evolution (see also \cite{Kirschner:1983di, Kirschner:1985cb,Kirschner:1994vc,Kirschner:1994rq,Griffiths:1999dj,Itakura:2003jp,Bartels:2003dj}); its resummation is usually referred to as the double logarithmic approximation (DLA). Here we will apply this method to determination of small-$x$ asymptotics for the valence quark transversity TMD in the same DLA limit. 

The paper is structured as follows. In Sec.~\ref{sec:trWil} we will derive an explicit expression for the quark transversely-polarized ``Wilson line". In Sec.~\ref{sec:Qtr} we will derive a relation between the quark transversity TMD at small-$x$ and the transversely-polarized dipole operator. The evolution equation for the transversely-polarized flavor non-singlet dipole operator is derived in Sec.~\ref{sec:evolA} and solved in Sec.~\ref{sec:sol}, leading to the small-$x$ asymptotics of valence quark transversity given in \eq{trans_asym} in the large-$N_c$ limit and in the DLA. The result \eqref{trans_asym} is consistent with one of the two possibilities for small-$x$ asymptotics of the chiral-odd PDF $h_1 (x, Q^2)$ found in \cite{Kirschner:1996jj}. The second possibility found in \cite{Kirschner:1996jj} is $h_1 (x, Q^2) \sim (1/x)^0$, which resulted from resumming the leading logarithms in $x$ (LLA), that is, powers of $\as \, \ln (1/x)$: such terms are outside of the DLA precision employed here. At the moment it appears that LLA kernel should enter the evolution equation(s) as an additive correction to the DLA kernel. Solution of the evolution equations with the leading and subleading kernels may result in the LLA term giving a multiplicative correction to our DLA small-$x$ asymptotics (see \cite{Chirilli:2013kca} for an example and further subtleties in the unpolarized evolution case), rather than generating an additive term suggested in \cite{Kirschner:1996jj}, though a more detailed investigation is needed to clarify this issue. This investigation is left for future work.


\section{Transversely polarized Wilson lines}
\label{sec:trWil}

\begin{figure}[ht]
\begin{center}
\includegraphics[width=  \textwidth]{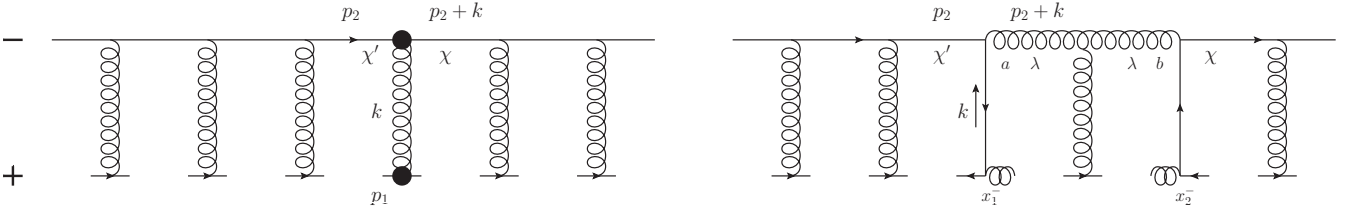} 
\caption{Transversely polarized fundamental Wilson line in the
  quasi-classical approximation in $\pd_\mu A^\mu =0$ Feynman gauge or in $A^-=0$ gauge. The black circles
  denote the spin-dependent sub-sub-eikonal scattering.}
\label{vpolT}
\end{center}
\end{figure}

In order to get a better understanding of the transversity observable in question we begin by constructing the transversely-polarized Wilson line operator for quarks. This operator is defined as transverse spin-dependent part of the scattering amplitude for a high-energy quark on a target. The target generates quasi-classical background quark and gluon fields. By analogy to helicity case, the leading contributions to the transversely-polarized Wilson line are given by the two diagrams shown in \fig{vpolT}. There, the black circles denote the sub-eikonal transverse spin-dependent part of the quark-gluon vertex. The high-energy quark at the top of each diagram is moving in the light-cone ``-" direction, while the target proton is moving in the ``+" direction. (For a 4-vector $v^\mu$ we define $v^\pm = (v^0 \pm v^3)/\sqrt{2}$.)

To work with the ``-" direction-moving quark we introduce the $+ \leftrightarrow -$ interchanged Brodsky-Lepage (BL) spinors \cite{Lepage:1980fj}, which we will also refer to as the anti-BL spinors:
\begin{align}
u_\sigma (p) = \frac{1}{\sqrt{\sqrt{2} \, p^-}} \, [\sqrt{2} \, p^- + m \, \gamma^0 +  \gamma^0 \, {\un \gamma} \cdot {\un p} ] \,  \rho (\sigma), \ \ \ v_\sigma (p) = \frac{1}{\sqrt{\sqrt{2} \, p^-}} \, [\sqrt{2} \, p^- - m \, \gamma^0 +  \gamma^0 \, {\un \gamma} \cdot {\un p} ] \,  \rho (-\sigma),
\end{align}
where $p^\mu = \left( \frac{{\un p}^2+ m^2}{2 p^-}, p^-, {\un p} \right)$, the transverse vectors are denotes by ${\un p} = (p^x, p^y)$ and
\begin{align}
  \rho (+1) \, = \, \frac{1}{\sqrt{2}} \, \left(
  \begin{array}{c}
      1 \\ 0 \\ -1 \\ 0
  \end{array}
\right), \ \ \ \rho (-1) \, = \, \frac{1}{\sqrt{2}} \, \left(
  \begin{array}{c}
        0 \\ 1 \\ 0 \\ 1
  \end{array}
\right) .
\end{align}
These spinors are eigenstates of $W^-$ such that $W^- u_\pm = \mp (p^-/2) \, u_\pm$, $W^- v_\pm = \pm (p^-/2) \, v_\pm$, with the Pauli-Lubanski vector $W_\mu = - \thalf \, \epsilon_{\mu\nu\rho\sigma} \, S^{\nu\rho} \, p^\sigma$ and $S^{\nu\rho} = (i/4) \, [\gamma^\nu , \gamma^\rho]$. 

To work with transverse spin it is convenient to define the transverse spinors which are eigenstates of $W_1$ \cite{Kovchegov:2012ga},
\begin{align}\label{chi_def}
u_\chi \equiv \frac{1}{\sqrt{2}} \, \left[ u_+ + \chi \, u_- \right], \ \ \ v_\chi \equiv \frac{1}{\sqrt{2}} \, \left[ v_+ + \chi \, v_- \right],
\end{align}
such that $W_1 u_\chi = (m/2) \chi u_\chi$, $W_1 v_\chi = - (m/2) \chi v_\chi$, but only for ${\un p} =0$. Hence these spinors are transverse to the quark momentum only for ${\un p} =0$.

Before we start our calculation, let us also note that in helicity basis the transverse spin is given by the helicity-flipping $({\un \tau})_{\sigma \sigma'}$ transverse Pauli matrices, while in the transverse spinor basis \eqref{chi_def} it is given by $(\tau_3)_{\chi \chi'}$, that is, by the $\chi \, \delta_{\chi \chi'}$ terms. Hence we are interested in the part of the quark scattering amplitude proportional to $\chi \, \delta_{\chi \chi'}$. To help us in the calculation, we obtain the following leading transverse polarization-dependent high-energy contributions to Dirac matrix elements of anti-BL spinors (by keeping the $\chi \, \delta_{\chi \chi'}$ terms only):
\begin{subequations}
\begin{align}
{\bar u}_\chi (p+k) \, \gamma^+ \, u_{\chi'} (p) & \supset - i \, m \, \chi \, \frac{1}{p^-} \, \delta_{\chi \chi'} \, {\un k} \times {\un S} \\
{\bar u}_\chi (p+k) \, \gamma^i \, u_{\chi'} (p) & \supset - i \,  m \, \chi \, \frac{k^-}{p^-} \, \delta_{\chi \chi'} \, \epsilon^{ij} S^j .
\end{align}
\end{subequations}

The transverse polarization-dependent gluon field of a plus-direction moving quark in $\pd_\mu A^\mu =0$ Feynman gauge is then\footnote{The gluon fields are indeed gauge-dependent. At this sub-eikonal level and beyond the Feynman gauge is not equivalent to the $A^-=0$ gauge. However, the operator we obtain this way is gauge-invariant (or, gauge-covariant) and is the same for all gauges in which the gluon fields do not stretch out to $x^-$-infinities and the gauge links at infinities never contribute. }
\begin{subequations}\label{fields}
\begin{align}
A^{a \, -} (x)  & = -  \frac{g}{2 \pi} \, t^a \, \frac{m \, \chi}{(p^+)^2} \, \delta (x^-) \, \frac{1}{x_\perp^2} 
\, \delta_{\chi \chi'} \, {\un x} \times {\un S} \\ 
A^{a \, i} (x)  & = -  \frac{g}{4 \pi} \, t^a \, \frac{m \, \chi}{(p^+)^2} \, \partial^+ \delta (x^-) \, \ln \frac{1}{x_\perp \, \Lambda} \, \delta_{\chi \chi'} \, \epsilon^{ij} S^j 
. \label{Atr}
\end{align}
\end{subequations}
Note the suppression by two powers of $p^+$ as compared to the eikonal field: thus the transverse polarization $\chi \, \delta_{\chi \chi'}$ dependence is sub-sub-eikonal. It also always comes in with the quark mass $m$, in agreement with the conventional wisdom. Here $\un S$ is a unit vector in the direction of the proton spin: in \eq{chi_def} we had ${\un S} = {\hat x}$.

As we mentioned already, the two possible contributions to the transversely polarized quark Wilson line are shown in \fig{vpolT} (see Appendix~\ref{A} for a discussion of other possible sub-sub-eikonal diagrams which do not contribute). We begin by calculating the contribution of the left panel in $\pd_\mu A^\mu =0$ gauge, neglecting multiple eikonal gluon exchanges first. The non-eikonal gluon exchange gives
\begin{align}
\frac{1}{2 p_2^-} \, i g \, {\bar u}_\chi (p_2+k) \, \slashed{A} (k)  \, u_{\chi'} (p_2) & = \frac{ i g \, m \, \chi}{2 (p_2^-)^2} \, \delta_{\chi \chi'} \, \left[  - i \,  {\un k} \times {\un S} \, A^- (k) + i \, k^- \,  {\un A} \times {\un S}   \right] \notag \\ & \approx  \frac{g \, m \, \chi}{2 \, (p_2^-)^2} \, \delta_{\chi \chi'} \, {\un k} \times {\un S}  \, A^- (k) ,
\end{align}
where we only keep the $\chi \, \delta_{\chi \chi'}$ terms. In the last step we have used the fact that $k^- \sim 1/p_1^+$ is negligibly small, constituting a sub-sub-sub-eikonal correction to the quantity of interest. Now $m$ is the mass of the projectile quark at the top of \fig{vpolT}.

Fourier transforming into coordinate space we get
\begin{align}
\chi \, \delta_{\chi, \chi'}  \, {\hat {\cal O}}^G_{pol, T} (x^-, {\un x}) \equiv & \int \frac{d k^+ \, d^2 k_\perp}{(2 \pi)^3} \, e^{- i k^+ x^- + i {\un k} \cdot {\un x}} \ \frac{ i g \, m \, \chi}{2 \, (p_2^-)^2} \, \left[ - i \, \delta_{\chi \chi'} \, {\un k} \times {\un S} \right] \, A^- (k) \notag \\ & = \frac{ g \, m \, \chi}{2 \, (p_2^-)^2} \, i \, \delta_{\chi \chi'} \, {\un S} \times {\un \nabla} \, A^- (x).
\end{align}
Inserting Wilson lines to account for the multiple eikonal gluon exchanges, we conclude that the gluon contribution to the transversely polarized quark Wilson line is
\begin{align} 
  \label{VpolT1}
  (V_{\ul x}^{pol, T})^G = \frac{2 g \, m \, (p_1^+)^2}{s^2} \, 
  \int\limits_{-\infty}^{+\infty} dx^- \: V_{\ul{x}}[+\infty, x^-] \:
   \left[ i \, {\un S} \times {\un \nabla}  \, A^-  (x^+ =0 , x^- , \ul{x}) \right] \: V_{\ul{x}} [x^- , -\infty].
\end{align}
The fundamental Wilson lines are denoted by
\begin{align}\label{Vdef}
  V_{\un{x}} [b^-, a^-] = \mathcal{P} \exp \left[ i g
    \int\limits_{a^-}^{b^-} d x^- \, A^+ (x^+=0, x^-, {\un x})
  \right].
\end{align}
Noticing that $\partial^- A^i =0$ for $A^i$ from \eq{Atr}, and that $[A^i, A^-] \sim 1/(p_1^+)^4$ is very strongly suppressed at high energy, we can complete $\nabla_i A^- = - \partial^i A^- \to F^{-i}$ and rewrite \eq{VpolT1} in a gauge-covariant form,
\begin{align} 
  \label{VpolT2}
  (V_{\ul x}^{pol, T})^G = \frac{2 g \, m \, (p_1^+)^2}{s^2}  \,
  \int\limits_{-\infty}^{+\infty} dx^- \: V_{\ul{x}}[+\infty, x^-] \: S^i \,
   \left[ i \, \epsilon^{ij} \, F^{-j} (x^- , \ul{x}) \right] \, V_{\ul{x}} [x^- , -\infty].
\end{align}
One can show explicitly that a calculation in $A^- =0$ gauge leads to the same result. The difference with the helicity case \cite{Kovchegov:2017lsr} is that now the ``transversely polarized Wilson line" \eqref{VpolT2} is doubly-suppressed at high energy and is proportional to the quark mass $m$. 

We next move on to the right panel in \fig{vpolT}, switching to the $A^- =0$ gauge. The contribution of the first (left) $t$-channel quark exchange is
\begin{align}\label{op1}
\frac{1}{2 p_2^-} \, \epsilon_\lambda^{\mu \, *} (p_2+k) \, (i g) \, {\bar \psi} (k) t^a \gamma_\mu u_{\chi'} (p_2) = - \frac{i g \, \chi'}{\sqrt{2} \,  \sqrt{\sqrt{2} \, p_2^-}} \, {\bar \psi} (k) \, t^a  \Bigg[  \delta_{\lambda, -1} \, \rho (+1) + \chi' \, \delta_{\lambda, 1} \, \rho(-1)  \notag \\  - \frac{m}{\sqrt{2} \, p_2^-} \, \left( \chi' \, \delta_{\lambda,  1} \, \gamma^0 \, \rho (-1) + \delta_{\lambda, -1} \, \gamma^0 \, \rho (+1) \right) - \frac{{\un \epsilon}_\lambda^* \cdot {\un k}}{p_2^- + k^-} \, \left( \chi' \, \gamma^0 \, \rho (+1) + \gamma^0 \, \rho (-1) \right)  \Bigg], 
\end{align}
again using the anti-BL spinors and neglecting terms further suppressed by $1/p_2^-$. Gluon polarization 4-vector in the $A^- =0$ gauge is $\epsilon^\mu_\lambda (p_2 + k) = \left( \frac{{\un \epsilon}_\lambda \cdot ({\un p}_2 + {\un k})}{p_2^- + k^-}, 0, {\un \epsilon}_\lambda \right)$ in terms of the $(+,-,\perp)$ and with ${\un \epsilon}_\lambda = (-1/\sqrt{2}) \, (\lambda, i)$ \cite{Lepage:1980fj}.

Fourier-transforming \eq{op1} yields
\begin{align}\label{op2}
& - \frac{i g \, \chi'}{\sqrt{2} \,  \sqrt{\sqrt{2} \, p_2^-}} \, {\bar \psi} (x_1^-, {\un x}) \, t^a \Bigg[  \delta_{\lambda, -1} \, \rho (+1) + \chi' \, \delta_{\lambda, 1} \, \rho(-1)  - \frac{m}{\sqrt{2} \, p_2^-} \, \left( \chi' \, \delta_{\lambda,  1} \, \gamma^0 \, \rho (-1) + \delta_{\lambda, +1} \, \gamma^0 \, \rho (+1) \right) \notag \\ & + i \, \frac{{\un \epsilon}_\lambda^* \cdot \overleftarrow{\un \nabla}}{p_2^-} \, \left( \chi' \, \gamma^0 \, \rho (+1) + \gamma^0 \, \rho (-1) \right)  \Bigg] \equiv - \frac{i g}{\sqrt{2} \,  \sqrt{\sqrt{2} \, p_2^-}} \, {\bar \psi} (x_1^-, {\un x}) \, t^a \, \overleftarrow{M} (\lambda, \chi')
\end{align}
with the transverse derivative acting on the antiquark field to its left. Note that $k^- \to i \, \partial^- = i \, \partial_+$, which we put to zero since ${\bar \psi}$ is independent of $x^+$ in our standard eikonal approximation: again, powers of $k^-$ correspond to terms further suppressed at high energy. The second (right) $t$-channel quark exchange in \fig{vpolT} gives
\begin{align}\label{op3}
& - \frac{i g \, \chi}{\sqrt{2} \,  \sqrt{\sqrt{2} \, p_2^-}}  \, t^b  \Bigg[ \delta_{\lambda, -1} \, \rho^T (+1) \, \gamma^0 + \chi \, \delta_{\lambda, 1} \, \rho^T(-1) \, \gamma^0  - \frac{m}{\sqrt{2} \, p_2^-} \, \left( \chi \, \delta_{\lambda, 1} \, \rho^T(-1) + \delta_{\lambda, -1} \, \rho^T (+1) \right) \notag \\ & - i \, \frac{{\un \epsilon}_\lambda \cdot {\un \nabla}}{p_2^-} \, \left( \chi \,  \rho^T(+1) + \rho^T (-1) \right) \Bigg]  \,  {\psi} (x_2^-, {\un x}) = - \frac{i g}{\sqrt{2} \,  \sqrt{\sqrt{2} \, p_2^-}}  \, t^b \, {\overrightarrow M}^\dagger (\lambda, \chi) \, \gamma^0 \,  {\psi} (x_2^-, {\un x}). 
\end{align}
Combining Eqs.~\eqref{op2} and \eqref{op3} we write an expression for the operator, the $\chi$-dependent part of 
which would give us the quark contribution to the transversely polarized Wilson line:
\begin{align} 
  \label{Vpolq1} 
 \chi \, \delta_{\chi, \chi'}  \,  (V^{pol, T}_{\un x})^q \subset - \frac{g^2 \, p_1^+ }{\sqrt{2} \, s}
  \int\limits_{-\infty}^\infty d x_1^- \, \int\limits_{x_1^-}^\infty d x_2^- \, \sum_\lambda \, & V_{\ul x} [+\infty, x_2^-] \: t^b  \, \left[ {\overrightarrow M}^\dagger (\lambda, \chi) \, \gamma^0 \, {\psi} (x_2^-, {\un x})  \right] \, U_{\ul x}^{ba} [ x_2^-,  x_1^-] \notag \\ & \times \, \left[ {\bar \psi} (x_1^-, {\un x}) \, 
\overleftarrow{M} (\lambda, \chi') \right] \: t^a \,V_{\ul x} [x_1^- , -\infty] .
\end{align}
Here $U_{\ul x}^{ba}$ is the adjoint Wilson line, defined by analogy to \eq{Vdef}. Summing over $\lambda$ and expanding over $1/p_2^-$ to the lowest non-trivial order we arrive at
\begin{align} 
  \label{Vpolq2} 
& \chi \, \delta_{\chi, \chi'}  \,  (V^{pol, T}_{\un x})^q \subset  - \frac{g^2 \, p_1^+ }{2 \, s} 
  \int\limits_{-\infty}^\infty d x_1^- \, \int\limits_{x_1^-}^\infty d x_2^- \, V_{\ul x} [+\infty, x_2^-] \:  t^b   \,  {\psi}_\beta (x_2^-, {\un x})  \, U_{\ul x}^{ba} [ x_2^-,  x_1^-]  \,  \Bigg\{ \gamma^5 \, \gamma^+ \, \delta_{\chi, - \chi'} + \delta_{\chi, \chi'} \, \gamma^+  \\ & + \frac{p_1^+}{s} \left\{ - 2 m  (\delta_{\chi, \chi'} - i \, \delta_{\chi, -\chi'} \, \gamma^1 \, \gamma^2 ) - \chi' \, \left[ (i \overleftarrow{\pd_x} + \overleftarrow{\pd_y}) \, \frac{1-\gamma^5}{2} \, (1 + i \, \gamma^1 \, \gamma^2) + (i \pd_x + \pd_y) \, \frac{1-\gamma^5}{2} \, (1 - i \, \gamma^1 \, \gamma^2)\right]  \right. \notag \\ & + \chi \, \left[ (i \overleftarrow{\pd_x} - \overleftarrow{\pd_y}) \, \frac{1+\gamma^5}{2} \, (1 - i \, \gamma^1 \, \gamma^2) + (i \pd_x - \pd_y) \, \frac{1+ \gamma^5}{2} \, (1 + i \, \gamma^1 \, \gamma^2)\right]  \notag \\ & \left. +  (i \overleftarrow{\pd_x} - \overleftarrow{\pd_y}) \, \frac{1+\gamma^5}{2} \, (i \, \gamma^2 - \gamma^1) + (i \pd_x + \pd_y) \, \frac{1+ \gamma^5}{2} \, (\gamma^1 + i \,  \gamma^2) \right\} + {\cal O} \left( \frac{1}{s^2} \right)  \Bigg\}_{\alpha\beta}  \, {\bar \psi}_\alpha (x_1^-, {\un x}) \, t^a \: V_{\ul x} [x_1^- , -\infty] . \notag
\end{align}

Since
\begin{align}
\chi \, \chi' = \delta_{\chi, \chi'} - \delta_{\chi,  -\chi'}, \ \ \  \chi - \chi' = 2 \chi \, \delta_{\chi,  -\chi'}, \ \ \ \chi + \chi' = 2 \chi \, \delta_{\chi,  \chi'}
\end{align}
only the $\chi$ and $\chi'$ terms in \eq{Vpolq2} have explicit dependence on polarization. Hence we write
\begin{align} 
  \label{Vpolq3} 
& \chi \, \delta_{\chi, \chi'}  \,  (V^{pol, T}_{\un x})^q \subset - \frac{g^2 \, (p_1^+)^2 }{2 \, s^2} 
  \int\limits_{-\infty}^\infty d x_1^- \, \int\limits_{x_1^-}^\infty d x_2^- \, V_{\ul x} [+\infty, x_2^-] \:  t^b   \,  {\psi}_\beta (x_2^-, {\un x})  \, U_{\ul x}^{ba} [ x_2^-,  x_1^-]  \\ &  \times \,  \Bigg\{ - \chi' \, \left[ (i \overleftarrow{\pd_x} + \overleftarrow{\pd_y}) \, \frac{1-\gamma^5}{2} \, (1 + i \, \gamma^1 \, \gamma^2) + (i \pd_x + \pd_y) \, \frac{1-\gamma^5}{2} \, (1 - i \, \gamma^1 \, \gamma^2)\right]   \notag \\ & + \chi \, \left[ (i \overleftarrow{\pd_x} - \overleftarrow{\pd_y}) \, \frac{1+\gamma^5}{2} \, (1 - i \, \gamma^1 \, \gamma^2) + (i \pd_x - \pd_y) \, \frac{1+ \gamma^5}{2} \, (1 + i \, \gamma^1 \, \gamma^2)\right]   \Bigg\}_{\alpha\beta}  \, {\bar \psi}_\alpha (x_1^-, {\un x}) \, t^a \: V_{\ul x} [x_1^- , -\infty] . \notag
\end{align}

Equivalently, one can rewrite this as
\begin{align} 
  \label{Vpolq4} 
& \chi \, \delta_{\chi, \chi'}  \,  (V^{pol, T}_{\un x})^q \subset - \frac{g^2 \, (p_1^+)^2 }{2 \, s^2} \, \chi
  \int\limits_{-\infty}^\infty d x_1^- \, \int\limits_{x_1^-}^\infty d x_2^- \, V_{\ul x} [+\infty, x_2^-] \:  t^b   \,  {\psi}_\beta (x_2^-, {\un x})  \, U_{\ul x}^{ba} [ x_2^-,  x_1^-] \,  \Big\{ (\delta_{\chi, \chi'} \gamma^5 + \delta_{\chi, -\chi'})  \\ &  \times \, \left[ \left( i \, \overleftarrow{\pd_x} - \gamma^5 \, \overleftarrow{\pd_y} \right) \, (1 - i \, \gamma^5 \, \gamma^1 \, \gamma^2) + (i \pd_x - \gamma^5 \, \pd_y) \, (1 + i \, \gamma^5 \, \gamma^1 \, \gamma^2) \right] \Big\}_{\alpha\beta}  \, {\bar \psi}_\alpha (x_1^-, {\un x}) \, t^a \: V_{\ul x} [x_1^- , -\infty] , \notag
\end{align}
or, keeping only the $\chi \, \delta_{\chi, \chi'}$ terms, 
\begin{align} 
  \label{Vpolq5} 
& (V^{pol, T}_{\un x})^q = - \frac{g^2 \, (p_1^+)^2 }{2 \, s^2}
  \int\limits_{-\infty}^\infty d x_1^- \, \int\limits_{x_1^-}^\infty d x_2^- \, V_{\ul x} [+\infty, x_2^-] \:  t^b   \,  {\psi}_\beta (x_2^-, {\un x})  \, U_{\ul x}^{ba} [ x_2^-,  x_1^-]  \\ &  \times \, \left[ \left( i \, \gamma^5 \, \overleftarrow{\pd_x} - \overleftarrow{\pd_y} \right) \, (1 - i \, \gamma^5 \, \gamma^1 \, \gamma^2) + (i \, \gamma^5 \, \pd_x - \pd_y) \, (1 + i \, \gamma^5 \, \gamma^1 \, \gamma^2) \right]_{\alpha\beta}  \, {\bar \psi}_\alpha (x_1^-, {\un x}) \, t^a \: V_{\ul x} [x_1^- , -\infty] . \notag
\end{align}
This result can be rewritten as
\begin{align} 
  \label{Vpolq6} 
& (V^{pol, T}_{\un x})^q = - \frac{g^2 \, (p_1^+)^2 }{2 \, s^2}
  \int\limits_{-\infty}^\infty d x_1^- \, \int\limits_{x_1^-}^\infty d x_2^- \, V_{\ul x} [+\infty, x_2^-] \:  t^b   \,  {\psi}_\beta (x_2^-, {\un x})  \, U_{\ul x}^{ba} [ x_2^-,  x_1^-]  \\ &  \times \, \left[  \left( i \, \gamma^5 \, \overleftarrow{\pd_x} - \overleftarrow{\pd_y} \right) \, \gamma^+ \, \gamma^- + (i \, \gamma^5 \, \pd_x - \pd_y) \, \gamma^- \, \gamma^+ \right]_{\alpha\beta}  \, {\bar \psi}_\alpha (x_1^-, {\un x}) \, t^a \: V_{\ul x} [x_1^- , -\infty] . \notag
\end{align}

To make \eq{Vpolq6} gauge-invariant (or, more precisely, gauge-covariant) we can replace $\pd_i \to D_i$, with $D_i$ being the covariant derivative, since the transverse gluon field $A^i$ of the plus-moving target in $\pd_\mu A^\mu =0$ Feynman gauge or in $A^- =0$ gauge is further suppressed by a power of energy, giving a sub-sub-sub-eikonal contribution to \eq{Vpolq6}, which is outside the precision of our approximation. The covariant derivatives still act only on the corresponding spinors, with the argument of $A^i$ in each derivative being the same as that of the spinor the derivative is acting on.

Combining Eqs.~\eqref{VpolT2} and \eqref{Vpolq6} we arrive at the full expression for the transversely polarized Wilson line
\begin{align} 
  \label{VpolT3}
 & V_{\ul x}^{pol, T} = \, \frac{2 g \, m \, (p_1^+)^2}{s^2}  \,
  \int\limits_{-\infty}^{+\infty} dx^- \: V_{\ul{x}}[+\infty, x^-] \: S^i \,
   \left[ i \, \epsilon^{ij} \, F^{-j} (x^- , \ul{x})  \right] \, V_{\ul{x}} [x^- , -\infty] \\
   &  - \frac{g^2 \, (p_1^+)^2 }{2 \, s^2}
  \int\limits_{-\infty}^\infty d x_1^- \, \int\limits_{x_1^-}^\infty d x_2^- \, V_{\ul x} [+\infty, x_2^-] \:  t^b   \,  {\psi}_\beta (x_2^-, {\un x})  \, U_{\ul x}^{ba} [ x_2^-,  x_1^-] \, \bigg[ \left( i \, \gamma^5 \, {\un S} \cdot \overleftarrow{\un D} - {\un S} \times \overleftarrow{\un D} \right) \,  \gamma^+ \, \gamma^- \notag \\ &  + (i \, \gamma^5 \, {\un S} \cdot {\un D} - {\un S} \times {\un D}) \, \gamma^- \, \gamma^+ \bigg]_{\alpha\beta}  \, {\bar \psi}_\alpha (x_1^-, {\un x}) \, t^a \: V_{\ul x} [x_1^- , -\infty] . \notag
\end{align}
The $1/s^2$ suppression in \eq{VpolT3} indicates that dependence on the transverse spin comes in as the sub-sub-eikonal effect at high energies. This result will persist through our analysis, leading to transversity quark TMDs being suppressed by $x^2$ compared to unpolarized TMDs at small $x$. 

Other sub-sub-eikonal diagrams, in addition to those considered in \fig{vpolT}, are analyzed in Appendix~\ref{A}, with the conclusion that they do not generate transverse spin dependence and can be discarded. 

Since the mass term cannot give the DLA evolution, only the mass-independent term in \eq{VpolT3} will lead to the DLA evolution below. Hence the DLA evolution of the quark polarized Wilson line does not generate polarized gluon emissions. Therefore, in the DLA approximation we will not need the transversely polarized gluon Wilson line, and the quark one in \eq{VpolT3} will be sufficient to construct the evolution equations.


\section{Quark Transversity Operator}
\label{sec:Qtr}

Our goal here is to simplify the quark transversity operator at small $x$. We begin with the definition of quark transversity TMDs (see e.g. \cite{Meissner:2007rx}):
\begin{align} \label{e:tdef}
h_{1T}^q (x, k_T^2) + \frac{k_x^2}{M^2} \, h_{1T}^{\perp \, q} (x, k_T^2)  & = \frac{1}{(2\pi)^3} \, \half \sum_{S_x = \pm 1} S_T^j \: \int d^2 r \, dr^- \,
e^{i k \cdot r} \bra{p, S_T} \bar\psi(0) \, \mathcal{U}[0,r] \, \frac{i \sigma^{j+} \gamma^5}{2} \,
\psi(r) \ket{p, S_T}_{r^+ = 0} \notag \\ & = \frac{1}{(2\pi)^3} \, \half \sum_{S_x = \pm 1} S_T^j \: \int d^2 r \, dr^- \,
e^{i k \cdot r} \bra{p, S_T} \bar\psi(0) \, \mathcal{U}[0,r] \, \frac{\gamma^5 \, \gamma^{+} \, \gamma^j }{2} \,
\psi(r) \ket{p, S_T}_{r^+ = 0} \notag \\ & = \frac{1}{(2\pi)^3} \, \int d^2 r \, dr^- \,
e^{i k \cdot r} \bra{p, S_x =  +1} \bar\psi(0) \, \mathcal{U}[0,r] \, \frac{\gamma^5 \, \gamma^{+} \, \gamma^1 }{2} \,
\psi(r) \ket{p, S_x = +1}_{r^+ = 0} .
\end{align}
(Here $M$ is the proton mass.) Note that the quark transversity TMD is usually defined as
\begin{align}\label{e:tdef_standard}
h_{1T}^q (x, k_T^2) + \frac{k_T^2}{2 \, M^2} \, h_{1T}^{\perp \, q} (x, k_T^2),
\end{align}
whereas in \eq{e:tdef} we have only taken the $x$-projection of the quark fields correlator. However, in the end we will use \eq{e:tdef} to determine both $h_{1T}^q$ and $h_{1T}^{\perp \, q}$ at small $x$, which would allow one to reconstruct the transversity TMD in \eq{e:tdef_standard}.

We choose the forward-pointing gauge link $\mathcal{U}[0,r]$, as appropriate for semi-inclusive deep inelastic scattering (SIDIS) process. In $A^-=0$ gauge that we adopt from now on, we have 
\begin{align}\label{UVV}
\mathcal{U}[0,r] = V_{\ul 0} [0, \infty] \: V_{\ul r} [\infty , r^-] .
\end{align}

\begin{figure}[ht]
\begin{center}
\includegraphics[width= 0.45 \textwidth]{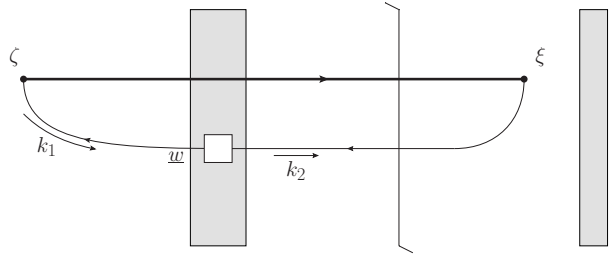} 
\caption{The diagram which, along with its complex conjugate, dominates the SIDIS cross section or, equivalently, quark TMDs at small $x$.}
\label{fig:B}
\end{center}
\end{figure}

To simplify \eq{e:tdef} we can either repeat the analysis from \cite{Kovchegov:2018znm} or redo the SIDIS cross section calculation from \cite{Kovchegov:2015zha}. In the end we end up with the diagram B from \cite{Kovchegov:2018znm} giving the leading small-$x$ contribution. The diagram is depicted in \fig{fig:B}, where shaded rectangles denote the target shock wave in the amplitude and in the complex conjugate amplitude, the square denotes the spin-dependent interaction with the polarized target, and the thick horizontal solid lines are the Wilson lines while the thin solid lines are antiquark propagators. Switching to the target-averaging notation employed in the saturation/Color Glass Condensate (CGC) \cite{Gribov:1984tu,Iancu:2003xm,Weigert:2005us,Jalilian-Marian:2005jf,Gelis:2010nm,Albacete:2014fwa,Kovchegov:2012mbw} approaches we write for the contribution of the diagram B from \fig{fig:B} (cf. Eq.~(10) in \cite{Kovchegov:2018znm})
\begin{align}\label{trTMD1}
h_{1T}^q (x, k_T^2)  + \frac{k_x^2}{M^2} \, h_{1T}^{\perp \, q} (x, k_T^2) = \frac{2 p^+}{(2\pi)^3} &  \:  \sum_{\bar{q}} \,  \int\limits_{-\infty}^0 d \zeta^- \, \int\limits^{\infty}_0 d \xi^- \, \int d^{2} \zeta \, d^{2} \xi 
\, e^{i k \cdot (\zeta - \xi)} \left( \thalf \gamma^5 \, \gamma^{+} \, \gamma^1 \right)_{\alpha \beta}
\notag \\ & \times \, \left\langle  \bar\psi_\alpha (\xi) \ket{\bar{q}} \: \bra{\bar{q}} \,
 V_{\ul \zeta} [\infty , -\infty] \,  \psi_\beta (\zeta) \right\rangle_{S_x =  +1} + \mbox{c.c.} .  
\end{align}
Repeating the calculation steps for the helicity TMD performed in \cite{Kovchegov:2018znm} we arrive at
\begin{align}\label{trTMD2}
& h_{1T}^q (x, k_T^2)  + \frac{k_x^2}{M^2} \, h_{1T}^{\perp \, q} (x, k_T^2)  = - \frac{2 p^+}{(2\pi)^3} \: \int d^{2} \zeta 
\, d^2 w \, \frac{d^2 k_1 \, d k_1^-}{(2\pi)^3} \, e^{i (\ul{k}_1 + {\un k}) \cdot (\ul{w} - \ul{\zeta})} \, \theta (k_1^-) \sum_{\chi_1, \, \chi_2} {\bar v}_{\chi_2} (k_2)  \thalf \gamma^5 \gamma^+ \gamma^1 v_{\chi_1} (k_1)  \\ & \times 
\, \Bigg\langle \mbox{T} \, V_{\ul \zeta}^{ij} [\infty , -\infty]  \ {\bar v}_{\chi_1} (k_1) \left( \hat{V}_{{\un w}}^\dagger \right)^{ji} v_{\chi_2} (k_2)  \Bigg\rangle_{S_x =  +1} \ \frac{1}{(2 x p^+ k_1^- + {\un k}_1^2)  \, (2 x p^+ k_1^- +  {\un k}^2) }  \Bigg|_{k_2^- = k_1^-, k_1^2 =0, k_2^2 =0, {\un k}_2 = - {\un k}}  + \mbox{c.c.} , \notag 
\end{align}
where now we chose the transverse basis for the spinors. By analogy to helicity operators we write 
\begin{align}\label{Vdecomp}
{\bar v}_{\chi} (p) \, \Big( \hat{V}_{\ul x}^\dagger \Big) \, v_{\chi'} (p') &=
2 \sqrt{p^- p^{\prime \, -}}\left( \delta_{\chi \chi'} \, V_{\ul x}^\dagger + \chi \, \delta_{\chi \chi'} \, V_{\ul x}^{pol, T \, \dagger} +  \ldots \right) .
\end{align}
The ellipsis denote the sub-eikonal and sub-sub-eikonal corrections (and beyond) independent of the proton transverse spin (but, in general, possibly dependent on $\chi$).

Employing anti-BL spinors we find
\begin{align}\label{mel}
{\bar v}_{\chi_2} (k_2)  \thalf \gamma^5 \gamma^+ \gamma^1 v_{\chi_1} (k_1) = \frac{1}{2 \, \sqrt{k_1^- \, k_2^-}} \, \Big[ & \, - \chi_1 \, \delta_{\chi_1 \chi_2} \, (2 \, {\un S} \cdot {\un k}_1 \, {\un S} \cdot {\un k}_2 - {\un k}_1 \cdot {\un k}_2 - m^2) + i \, m \,  \delta_{\chi_1 \chi_2} \, {\un S} \times ({\un k}_1 - {\un k}_2)  \\ & - i \, \chi_1 \, \delta_{\chi_1, \, - \chi_2} \, ({\un S} \times {\un k}_1 \, {\un S} \cdot {\un k}_2 + {\un S} \cdot {\un k}_1 \, {\un S} \times {\un k}_2 ) - m \,  \delta_{\chi_1, \, - \chi_2} \, {\un S} \cdot ({\un k}_1 + {\un k}_2) \Big] ,  \notag
\end{align}
where, again, ${\un S} = {\hat x}$ is the unit vector in the direction of the proton spin. 

Substituting Eqs.~\eqref{Vdecomp} and \eqref{mel} into \eq{trTMD2} we arrive at
\begin{align}\label{trTMD3}
& h_{1T}^q (x, k_T^2)  + \frac{k_x^2}{M^2} \, h_{1T}^{\perp \, q} (x, k_T^2)  =  \frac{4 p^+}{(2\pi)^3}   \:  \int d^{2} \zeta 
\, d^2 w \, \frac{d^2 k_1 \, d k_1^-}{(2\pi)^3} \, e^{i (\ul{k}_1 + {\un k}) \cdot (\ul{w} - \ul{\zeta})} \, \theta (k_1^-) \ \frac{1}{(2 x p^+ k_1^- + {\un k}_1^2)  \, (2 x p^+ k_1^- +  {\un k}^2) } \notag \\ & \times \, \Bigg\{ (- 2 \, {\un S} \cdot {\un k}_1 \, {\un S} \cdot {\un k} + {\un k}_1 \cdot {\un k} -m^2 )   
\, \left\langle \mbox{T} \, \tr \left[ V_{\ul \zeta} \, V_{\un w}^{pol, T \, \dagger} \right]  \right\rangle_{S_x =  +1}  + i \, m \, {\un S} \times ({\un k}_1 + {\un k}) \, \left\langle \mbox{T} \, \tr \left[ V_{\ul \zeta} \, V_{\un w}^{\dagger} \right]  \right\rangle_{S_x =  +1}  \Bigg\} + \mbox{c.c.} .   
\end{align}
Adding in the complex conjugate (where we interchange $\zeta \leftrightarrow w$ after conjugation) yields
\begin{align}\label{trTMD333}
& h_{1T}^q (x, k_T^2)  + \frac{k_x^2}{M^2} \, h_{1T}^{\perp \, q} (x, k_T^2)  = \frac{4 p^+}{(2\pi)^3}   \:  \int d^{2} \zeta 
\, d^2 w \, \frac{d^2 k_1 \, d k_1^-}{(2\pi)^3} \, e^{i (\ul{k}_1 + {\un k}) \cdot (\ul{w} - \ul{\zeta})} \, \theta (k_1^-) \ \frac{1}{(2 x p^+ k_1^- + {\un k}_1^2)  \, (2 x p^+ k_1^- +  {\un k}^2) } \notag \\ & \times \, \Bigg\{ \left( - 2 \, {\un S} \cdot {\un k}_1 \, {\un S} \cdot {\un k} + {\un k}_1 \cdot {\un k} -m^2 \right)  
\, \left\langle \mbox{T} \, \tr \left[ V_{\ul \zeta} \, V_{\un w}^{pol, T \, \dagger} \right] +  \overline{\mbox{T}} \, \tr \left[ V_{\un \zeta}^{pol, T} \, V_{\ul w}^\dagger \right]  \right\rangle_{S_x =  +1}  \\ & + i \, m \, {\un S} \times ({\un k}_1 + {\un k}) \,\left\langle  \mbox{T} \, \tr \left[ V_{\ul \zeta} \, V_{\un w}^{\dagger} \right] - \overline{\mbox{T}} \, \tr \left[ V_{\un \zeta} \, V_{\ul w}^{\dagger} \right]  \right\rangle_{S_x =  +1}  \Bigg\} .  \notag 
\end{align}
The last term in the brackets is zero \cite{Kovchegov:2018znm}, since
\begin{align}
\left\langle  \mbox{T} \, \tr \left[ V_{\ul \zeta} \, V_{\un w}^{\dagger} \right] - \overline{\mbox{T}} \, \tr \left[ V_{\un \zeta} \, V_{\ul w}^{\dagger} \right]  \right\rangle_{S_x =  +1} = 0
\end{align}
for true Wilson lines, if we flip the  Wilson lines in the second trace from the complex conjugate amplitude into the amplitude using the reflection symmetry employed in \cite{Kovchegov:2018znm}, which was verified in \cite{Mueller:2012bn} up to next-to-leading logarithms (NLL) in $x$ for the unpolarized evolution. We are left with
\begin{align}\label{trTMD3333}
 h_{1T}^q (x, k_T^2)  + \frac{k_x^2}{M^2} \, h_{1T}^{\perp \, q} (x, k_T^2)  = & \, \frac{4 p^+}{(2\pi)^3}   \:  \int d^{2} \zeta 
\, d^2 w \, \frac{d^2 k_1 \, d k_1^-}{(2\pi)^3} \, e^{i (\ul{k}_1 + {\un k}) \cdot (\ul{w} - \ul{\zeta})} \, \theta (k_1^-)  \ \frac{1}{(2 x p^+ k_1^- + {\un k}_1^2)  \, (2 x p^+ k_1^- +  {\un k}^2) } \notag \\ & \times \, \left( - 2 \, {\un S} \cdot {\un k}_1 \, {\un S} \cdot {\un k} + {\un k}_1 \cdot {\un k} -m^2 \right)  
\, \left\langle \mbox{T} \, \tr \left[ V_{\ul \zeta} \, V_{\un w}^{pol, T \, \dagger} \right] +  \overline{\mbox{T}} \,\tr \left[ V_{\un \zeta}^{pol, T} \, V_{\ul w}^\dagger \right]  \right\rangle_{S_x =  +1} .
\end{align}

The flavor-singlet and non-singlet (valence quark) distributions are defined by
\begin{subequations}
\begin{align}
& h_{1T}^S (x, k_T^2) = \sum_f \, \left[ h_{1T}^{q_f} (x, k_T^2)  + h_{1T}^{{\bar q}_f} (x, k_T^2) \right], \ \ \  h_{1T}^{\perp \, S} (x, k_T^2) = \sum_f \, \left[ h_{1T}^{\perp \, q_f} (x, k_T^2) + h_{1T}^{\perp \, {\bar q}_f} (x, k_T^2) \right], \\
& h_{1T}^{NS} (x, k_T^2) = h_{1T}^q (x, k_T^2)  - h_{1T}^{\bar q} (x, k_T^2) , \ \ \  h_{1T}^{\perp \, NS} (x, k_T^2) = h_{1T}^{\perp \, q} (x, k_T^2) - h_{1T}^{\perp \, {\bar q}} (x, k_T^2) .
\end{align}
\end{subequations}

The singlet version of \eq{trTMD3333} is
\begin{align}\label{trTMD3334}
 & \, h_{1T}^S (x, k_T^2)  + \frac{k_x^2}{M^2} \, h_{1T}^{\perp \, S} (x, k_T^2)  = \frac{4 p^+}{(2\pi)^3}  \, \sum_f \:  \int d^{2} \zeta 
\, d^2 w \, \frac{d^2 k_1 \, d k_1^-}{(2\pi)^3} \, e^{i (\ul{k}_1 + {\un k}) \cdot (\ul{w} - \ul{\zeta})} \, \theta (k_1^-)  \ \frac{1}{(2 x p^+ k_1^- + {\un k}_1^2)  \, (2 x p^+ k_1^- +  {\un k}^2) } \notag \\ & \times \, \left( - 2 \, {\un S} \cdot {\un k}_1 \, {\un S} \cdot {\un k} + {\un k}_1 \cdot {\un k} -m_f^2 \right)  
\, \left\langle \mbox{T} \, \tr \left[ V_{\ul \zeta} \, V_{\un w}^{pol, T \, \dagger} \right] + \mbox{T} \, \tr \left[ V_{\un w}^{pol, T} \, V_{\ul \zeta}^\dagger \right] +  \overline{\mbox{T}} \,\tr \left[ V_{\un \zeta}^{pol, T} \, V_{\ul w}^\dagger \right] + \overline{\mbox{T}} \,\tr \left[ V_{\ul w} \, V_{\un \zeta}^{pol, T \, \dagger}  \right]  \right\rangle_{S_x =  +1} .
\end{align}

Define the doubly energy-rescaled transversely-polarized flavor-singlet dipole amplitude
\begin{align}\label{Tdef}
T_{10}^S (z s) & \equiv \frac{(z s)^2}{2 N_c} \, \mbox{Re} \, \left\langle \mbox{T} \, \tr \left[ V_{\ul 0} \, V_{\un 1}^{pol, T \, \dagger} \right] + \mbox{T} \, \tr \left[ V_{\un 1}^{pol, T} \, V_{\ul 0}^\dagger \right]  \right\rangle_{S_x = +1} \\ & = \frac{2 (p^+ k_1^-)^2}{N_c}  \, \mbox{Re} \, \left\langle \mbox{T} \, \tr \left[ V_{\ul 0} \, V_{\un 1}^{pol, T \, \dagger} \right] + \mbox{T} \,  \tr \left[ V_{\un 1}^{pol, T} \, V_{\ul 0}^\dagger \right]  \right\rangle_{S_x = +1} \notag
\end{align}
with $z s = 2 p^+ k_1^-$. Further note that the integrals like
\begin{align}\label{int_tr1}
\int d^2 \left( \frac{x_0 + x_1}{2} \right) \, \left\langle \mbox{T} \, \tr \left[ V_{\ul 0} \, V_{\un 1}^{pol, T \, \dagger} \right] \right\rangle_{S_x = +1}
\end{align}
(for any of the correlators in \eq{trTMD3334}) are even functions of ${\un x}_{10} = {\un x}_1 - {\un x}_0$. The reason for this is slightly different from the helicity case: in case of helicity we could argue that the integrals like \eqref{int_tr1} are simply functions of $|{\un x}_1 - {\un x}_0|$ due to the absence of a preferred transverse direction in the problem. Now that the target proton and the polarized projectile quark in the dipole both have transverse spins, ${\un S}_T$ and ${\un S}_P$. Since PT symmetry requires that spin dependence enters as a term bilinear in the two spins, we write
 \begin{align}\label{int_tr2}
\int d^2 \left( \frac{x_0 + x_1}{2} \right) \, \left\langle \mbox{T} \, \tr \left[ V_{\ul 0} \, V_{\un 1}^{pol, T \, \dagger} \right] \right\rangle_{S_T} = A (x_{10}) + B (x_{10}) \, {\un S}_T \cdot {\un x}_{10} \,  {\un S}_P \cdot {\un x}_{10} + \ldots ,
\end{align}
where $A$ and $B$ are some scalar functions of $x_{10} = |{\un x}_1 - {\un x}_0|$ while the ellipsis denote terms like $B$ but with either or both scalar products replaced by a vector product of the same vectors. We thus see that the object on the left of \eq{int_tr2} (along with impact-parameter integrals of other such correlators in \eq{trTMD3334}) is an even function under the ${\un x}_1 \leftrightarrow {\un x}_0$ interchange.

Employing this observation along with the definition \eqref{Tdef} in \eq{trTMD3334} yields for the flavor-singlet case
\begin{align}\label{trTMD44}
& \, h_{1T}^S (x, k_T^2)  + \frac{k_x^2}{M^2} \, h_{1T}^{\perp \, S} (x, k_T^2)  =  \frac{8 N_c}{(2\pi)^3}   \: \sum_f \,  \int d^{2} x_0 
\, d^2 x_1 \, \frac{d^2 k_1 \, d k_1^-}{2 p^+ \, (k_1^-)^2 \, (2\pi)^3} \, e^{i (\ul{k}_1 + {\un k}) \cdot {\un x}_{10}} \, \theta (k_1^-) \notag \\ & \times \, \frac{1}{(2 x p^+ k_1^- + {\un k}_1^2)  \, (2 x p^+ k_1^- +  {\un k}^2) }  \, \left( - 2 \, {\un S} \cdot {\un k}_1 \, {\un S} \cdot {\un k} + {\un k}_1 \cdot {\un k} -m_f^2 \right) \, T^S_{10} (z s)  .
\end{align}

For the flavor non-singlet case, we define the transversely-polarized flavor non-singlet dipole amplitude
\begin{align}\label{TNSdef}
T^{NS}_{10} (z s) & \equiv \frac{(z s)^2}{2 N_c} \, \mbox{Re} \, \left\langle \mbox{T} \, \tr \left[ V_{\ul 0} \, V_{\un 1}^{pol, T \, \dagger} \right] - \mbox{T} \, \tr \left[ V_{\un 1}^{pol, T} \, V_{\ul 0}^\dagger \right]  \right\rangle_{S_x = +1} \\ & = \frac{2 (p^+ k_1^-)^2}{N_c} \, \mbox{Re} \,  \left\langle \mbox{T} \, \tr \left[ V_{\ul 0} \, V_{\un 1}^{pol, T \, \dagger} \right] - \mbox{T} \,  \tr \left[ V_{\un 1}^{pol, T} \, V_{\ul 0}^\dagger \right]  \right\rangle_{S_x = +1} \notag
\end{align}
which gives, for the valence-quark transversity TMDs,
\begin{align}\label{trTMD_NS}
& \, h_{1T}^{NS} (x, k_T^2)  + \frac{k_x^2}{M^2} \, h_{1T}^{\perp \, NS} (x, k_T^2)  =  \frac{8 N_c}{(2\pi)^3}   \:  \int d^{2} x_0 
\, d^2 x_1 \, \frac{d^2 k_1 \, d k_1^-}{2 p^+ \, (k_1^-)^2 \, (2\pi)^3} \, e^{i (\ul{k}_1 + {\un k}) \cdot {\un x}_{10}} \, \theta (k_1^-) \notag \\ & \times \, \frac{1}{(2 x p^+ k_1^- + {\un k}_1^2)  \, (2 x p^+ k_1^- +  {\un k}^2) } \, \left( - 2 \, {\un S} \cdot {\un k}_1 \, {\un S} \cdot {\un k} + {\un k}_1 \cdot {\un k} -m^2 \right) \, T^{NS}_{10} (z s) .
\end{align}

Continuing with the non-singlet distribution, if we simply put $x \to 0$ in \eq{trTMD_NS}, we get a non-logarithmic (in $z$) but non-zero contribution: integrating over ${\un k}_1$ we arrive at
\begin{align}\label{trTMD7}
h_{1T}^{NS} (x, k_T^2)  + \frac{k_x^2}{M^2} \, h_{1T}^{\perp \, NS} (x, k_T^2)  = & \,  \frac{8 N_c}{(2\pi)^5 \, k_\perp^2}   \: \int d^{2} x_0 
\, d^2 x_1 \, e^{i {\un k} \cdot \ul{x}_{10}} \, \frac{1}{s} \int\limits_{\Lambda^2/s}^1 \frac{d z}{z^2}   \\ & \times \left[- 2 \, i \, \frac{{\un S} \cdot \ul{x}_{10}}{x_{10}^2} \, {\un S} \cdot {\un k} + i \, \frac{\ul{x}_{10} \cdot {\un k}}{x_{10}^2} -m^2 \, \ln \frac{1}{x_{10} \, \sqrt{x z s}}  \right]  \notag 
\, T^{NS}_{10} (z s) .  \notag
\end{align}
If $T^{NS}_{10} (z s, {\un S})$ obeys a regular DLA-type evolution, that is if
\begin{align}\label{sample_asym}
T^{NS}_{10} (z s) \sim (z s)^{C \sqrt{\as}}
\end{align}
with $C$ a constant, then the $z$-integral in \eq{trTMD7} would be dominated by its lower limit (for $\as \ll 1$), leading to 
\begin{align}\label{trTMD??}
h_{1T}^{NS} (x, k_T^2) \sim h_{1T}^{\perp \, NS} (x, k_T^2) \sim \left( \frac{1}{x} \right)^0 .
\end{align}
The same conclusion would be valid for any perturbative positive power in \eq{sample_asym}. It appears that the asymptotics \eqref{trTMD??} agrees with that found in in \cite{Kirschner:1996jj}.

However, the $dz/z^2$ integral in \eq{trTMD7} is dominated by the lower limit of the $z$-integral, $z \sim \Lambda^2/s$. In this limit the antiquarks $k_1$ and $k_2$ in \fig{fig:B} do not live a long enough time to get outside the shock wave, as the figure suggest. (The antiquark lifetime is $2 z p_2^-/\perp^2 = \Lambda^2/(p_1^+ \perp^2) \sim 1/p_1^+$, which is exactly the shock wave width.) We conclude that the asymptotics \eqref{trTMD??} originates from the contribution where the diagram B is marginally applicable. To understand it better, we need to review the diagram where both vertices $\zeta, \xi$ are inside the shock wave.

\begin{figure}[ht]
\begin{center}
\includegraphics[width= 0.4 \textwidth]{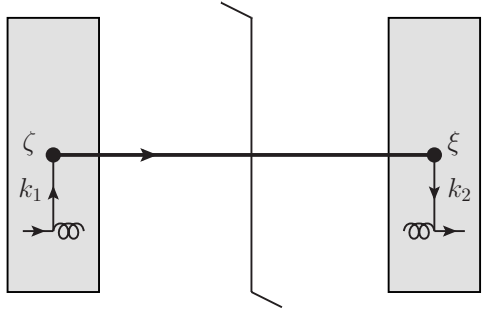} 
\caption{The diagram where both Wilson lines originate inside the shock wave on the different sides of the cut.}
\label{fig:A}
\end{center}
\end{figure}

The contribution with $\zeta, \xi$ inside the shock wave was captured by the diagram A in the notation of \cite{Kovchegov:2018znm}. It is depicted in \fig{fig:A} here, with a sample interaction with the shock waves shown explicitly too. In \cite{Kovchegov:2018znm} we have shown that all the radiative corrections to this diagram are canceled (in DLA) by other diagrams. The contribution of the diagram \fig{fig:A} without radiative corrections is constant in $x$. In the case of helicity TMDs, a constant was subleading compared to the leading small-$x$ helicity asymptotics, and was discarded. Now we see that a constant in $x$ behavior is of the same order as \eq{trTMD??}. Therefore, diagram A has to be included in our analysis. 

The constant in $x$ part of the diagram A in \fig{fig:A} comes from the instantaneous terms \cite{Lepage:1980fj,Brodsky:1997de} in the $k_1$ and $k_2$ propagators. To include this contribution of diagram A, we simply have to include (or, more precisely, insert back) the instantaneous terms into those propagators in the diagram B. One can show that the instantaneous contributions, when added to the diagram B, replace
\begin{align}\label{inst1}
\slashed{k}_1 \, \theta (k_1^-) \, \frac{1}{2 k^+ k_1^- + {\un k}_1^2} \to \slashed{k}_1 \, \theta (k_1^-) \, \frac{1}{2 k^+ k_1^- + {\un k}_1^2} - \frac{1}{2 k_1^-} \, \gamma^-
\end{align} 
(and ditto for $k_2$) after one of the steps omitted in arriving at \eq{trTMD2} if we now understand the $\zeta^-$ and $\xi^-$ integrals in \eq{trTMD1} as {\sl including} the points $\zeta^-=0$ and $\xi^-=0$, such that we could pick up the delta-functions there. (These are the contributions coming from the diagram A.) Here $\slashed{k_1} = \gamma^- \frac{{\un k}_1^2}{2 k_1^-} + \gamma^+ k_1^- - {\un \gamma} \cdot {\un k}_1$. At small $k_1^-$ we have $\slashed{k_1} \approx \gamma^- \frac{{\un k}_1^2}{2 k_1^-}$ such that (for $k_1^- >0$)
\begin{align}
\left[ \slashed{k}_1 \, \theta (k_1^-) \, \frac{1}{2 k^+ k_1^- + {\un k}_1^2} - \frac{1}{2 k_1^-} \, \gamma^- \right] \Bigg|_{\mbox{small} \ k_1^-} \approx \slashed{k}_1 \, \frac{-2 k^+ k_1^-}{{\un k}_1^2 \, (2 k^+ k_1^- + {\un k}_1^2)}.
\end{align}
Therefore, the instantaneous terms regulate the $k_1^-$ integral by a lower cutoff at $k_1^- \approx \frac{{\un k}_1^2}{2 k^+}$. This means that \eq{trTMD7} now contains a new lower bound on the $z$-integral and is proportional to 
\begin{align}
\frac{1}{s} \int\limits_{{\un k}_1^2/(x \, s)}^1  \frac{d z}{z^2} = \frac{x}{{\un k}_1^2} - \frac{1}{s} = \frac{x}{{\un k}_1^2} - \frac{x}{Q^2} = {\cal O} (x). 
\end{align}

Indeed \eq{inst1} also contains a contribution of the instantaneous term for $k_1^- <0$. This part would regulate the $k_1^-$-integral in the ``c.c." part of \eq{trTMD2}. Simply including $\zeta^-=0$ and $\xi^-=0$ contributions into both integrals in both terms of \eq{trTMD1} would introduce double-counting: we have to include those points only once, and then split the contribution of the resulting instantaneous term between the two terms (time-orderings). 

We conclude that the contribution of \eq{trTMD7} is smaller than it seems naively, being $\sim x$ rather than a constant in $x$. However, at this order other terms in the full \eq{trTMD_NS} may contribute as well. If we expand \eq{trTMD_NS} to order-$x$, we obtain
\begin{align}\label{trTMD8NS}
 h_{1T}^{NS} (x, k_T^2)  + \frac{k_x^2}{M^2} \, h_{1T}^{\perp \, NS} (x, k_T^2) = & \, - x \, \frac{8 N_c}{(2\pi)^4}   \: \int d^{2} x_0 
\, d^2 x_1 \, \int\limits_{\Lambda^2/s}^1 \frac{d z}{z} \int \frac{d^2 k_1 }{(2\pi)^2} \, e^{i (\ul{k}_1 + {\un k}) \cdot {\un x}_{10}} \, \frac{1}{{\un k}_1^2 \, {\un k}^2} \left[ \frac{1}{{\un k}_1^2} +  \frac{1}{{\un k}^2}
\right]  \notag \\ & \times \, \left( - 2 \, {\un S} \cdot {\un k}_1 \, {\un S} \cdot {\un k} + {\un k}_1 \cdot {\un k} -m^2 \right)  
\, T^{NS}_{10} (z s) .
\end{align}
This contribution comes in with a logarithmic integral in $z$. For the asymptotics of $T^{NS}_{10} (z s)$ from \eq{sample_asym} it leads to
\begin{align}\label{trTMD_DLA}
h_{1T}^{NS} (x, k_T^2) \sim h_{1T}^{\perp \, NS} (x, k_T^2) \sim \left( \frac{1}{x} \right)^{-1 + C \, \sqrt{\as} } ,
\end{align}
which, for $C>0$, would dominate over $\sim x$ contribution of \eq{trTMD7}. We tentatively conclude that \eq{trTMD8NS} may give us the true small-$x$ asymptotics of the flavor non-singlet quark transversity TMD. 

Repeating the above steps for the flavor-singlet case we arrive at
\begin{align}\label{trTMD8}
 h_{1T}^S (x, k_T^2)  + \frac{k_x^2}{M^2} \, h_{1T}^{\perp \, S} (x, k_T^2) = & \, - x \, \frac{8 N_c}{(2\pi)^4}   \: \sum_f \,  \int d^{2} x_0 
\, d^2 x_1 \, \int\limits_{\Lambda^2/s}^1 \frac{d z}{z} \int \frac{d^2 k_1 }{(2\pi)^2} \, e^{i (\ul{k}_1 + {\un k}) \cdot {\un x}_{10}} \, \frac{1}{{\un k}_1^2 \, {\un k}^2} \left[ \frac{1}{{\un k}_1^2} +  \frac{1}{{\un k}^2}
\right]  \notag \\ & \times \, \left( - 2 \, {\un S} \cdot {\un k}_1 \, {\un S} \cdot {\un k} + {\un k}_1 \cdot {\un k} -m_f^2 \right)  
\, T^S_{10} (z s)  .
\end{align}


\section{Evolution of the flavor non-singlet transversely polarized dipole}
\label{sec:evol}


\subsection{Operator evolution}
\label{sec:evolA}

The discussion here is for the flavor non-singlet (valence-quark) distribution. We now want to construct the evolution for the transversely polarized dipole amplitude
\begin{align}
T_{10}^{NS} (z s) = & \frac{(z \, s)^2}{2 N_c} \, \mbox{Re} \: \left\langle \mbox{T} \, \tr \left[ V_{\ul 0} \, V_{\un 1}^{pol, T \, \dagger} \right] - \mbox{T} \,  \tr \left[ V_{\un 1}^{pol, T} \, V_{\ul 0}^\dagger \right]  \right\rangle_{S_x = +1} .
\end{align}
Quark mass cannot contribute to the DLA evolution: therefore, polarized gluon emissions do not contribute. In addition, by analogy to the flavor non-singlet helicity case \cite{Kovchegov:2016zex}, eikonal gluon exchanges do not contribute to $T_{10}^{NS} (z s)$ either. We are left only with polarized soft quark emissions contribution to the small-$x$ evolution. Therefore we discard the gluon-exchange term in \eq{VpolT3} and write
\begin{align}\label{T1}
& T^{NS}_{10} (z s)  = - \frac{g^2 \, (p^+)^2 }{4 \, N_c}   \int\limits_{-\infty}^\infty d x_1^- \, \int\limits_{x_1^-}^\infty d x_2^- \, \mbox{Re} \, \Bigg\langle \mbox{T} \, \tr \Bigg[ V_{\ul 0}^\dagger \, V_{\ul 1} [+\infty, x_2^-] \:  t^b   \,  {\psi}_\beta (x_2^-, {\un x}_1)  \, U_{{\ul 1}}^{ba} [ x_2^-,  x_1^-]  \\ & \times \, \bigg[  \left( i \, \gamma^5 \, {\un S} \cdot \overleftarrow{\un D}_1 - {\un S} \times \overleftarrow{\un D}_1 \right) \,  \gamma^+ \, \gamma^-   + (i \, \gamma^5 \, {\un S} \cdot {\un D}_1 - {\un S} \times {\un D}_1) \, \gamma^- \, \gamma^+ \bigg]_{\alpha\beta}  \, {\bar \psi}_\alpha (x_1^-, {\un x}_1) \, t^a \: V_{{\ul 1}} [x_1^- , -\infty] \Bigg]  -  \mbox{c.c.}  \Bigg\rangle_{S_x = +1}  . \notag
\end{align}
(Note that the operator in the angle brackets is real, but not hermitean.) As usual, the derivatives act only on the spinors and the subscript $1$ refers to them acting on ${\un x}_1$. Simplifying \eq{T1} in the large-$N_c$ linearized approximation, where all fundamental traces made out of only true Wilson lines are replaced by $N_c$, along with replacing covariant derivatives by the partial derivatives, we arrive at (note the sign change when moving $\psi$ past $\bar \psi$)
\begin{align}\label{T2}
T^{NS}_{10} (z s) =  \frac{g^2 \, (p^+)^2 }{8} \, \int\limits_{-\infty}^\infty d x_1^- \, \int\limits_{x_1^-}^\infty d x_2^- \, & \mbox{Re} \, \Bigg\langle \mbox{T} \, \tr \Bigg[ {\bar \psi} (x_1^-, {\un x}_1) \, \bigg[  \left( i \, \gamma^5 \, \pd_x^1 - \pd_y^1 \right) \,  \gamma^+ \, \gamma^-  \\ &  + (i \, \gamma^5 \, \overleftarrow{\pd}_x^1 - \overleftarrow{\pd}_y^1) \, \gamma^- \, \gamma^+ \bigg]  \, V_{\ul 1} [x_1^-, x_2^-] \,  {\psi} (x_2^-, {\un x}_1)  \Bigg] -  \mbox{c.c.}  \Bigg\rangle_{S_x = +1}  . \notag
\end{align}

\begin{figure}[ht]
\begin{center}
\includegraphics[width= 0.7 \textwidth]{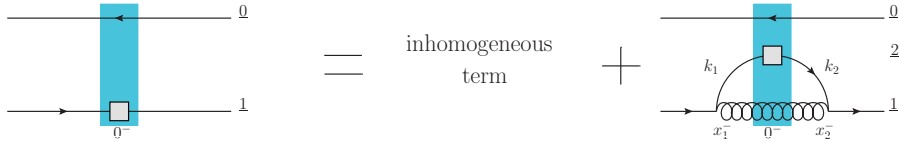} 
\caption{Small-$x$ evolution of the transversely polarized dipole amplitude.}
\label{fig:evol}
\end{center}
\end{figure}

One step of the evolution, as pictured in \fig{fig:evol}, leads to
\begin{align}\label{Tev1}
& \delta T_{10}^{NS} (z s) =  - \, \frac{g^2 \, (p^+)^2 }{8}  \int\limits_{-\infty}^0 d x_1^- \, \int\limits_{0}^\infty d x_2^- \, \int d^2 x_2 \, \frac{d^4 k_1}{(2 \pi)^4} \, \frac{d^4 k_2}{(2 \pi)^4} \, e^{i k_1^+ x_1^- + i {\un k}_1 \cdot {\un x}_{21}} \, e^{- i k_2^+ x_2^- + i {\un k}_2 \cdot {\un x}_{12}} \, \mbox{Re} \, \Bigg\langle \mbox{T} \, i \, (V_{\ul 1}^\dagger)^{ji} \,  \\ & \times  \, \tr \Bigg[ \frac{i \slashed{k_1}}{k_1^2 + i \epsilon} \, \bigg[  -  \left( i \, \gamma^5 \, k_2^x - k_2^y \right) \,  \gamma^+ \, \gamma^-  + (i \, \gamma^5 \, k_1^x - k_1^y) \, \gamma^- \, \gamma^+ \bigg]  \,  \frac{i \slashed{k_2}}{k_2^2 + i \epsilon} \,  \left( {\hat V}_{\un 2} \right)^{ij}  \Bigg]  (2 \pi) \, \delta (k_1^- - k_2^-) - \mbox{c.c.} \Bigg\rangle_{S_x = +1}   . \notag
\end{align}
Integrating over $k_2^-$, $k_1^+$ and $k_2^+$ we get
\begin{align}\label{Tev2}
& \delta T_{10}^{NS} (z s) =  -  \, \frac{g^2 \, (p^+)^2 }{8}  \int\limits_{-\infty}^0 d x_1^- \, \int\limits_{0}^\infty d x_2^- \, \int d^2 x_2 \, \frac{d k_1^- \, d^2 k_1}{(2 \pi)^3 \, (2 k_1^-)^2} \, \frac{d^2 k_2}{(2 \pi)^2} \, e^{i \frac{{\un k}_1^2}{2 k_1^-} x_1^- + i {\un k}_1 \cdot {\un x}_{21}} \, e^{- i \frac{{\un k}_2^2}{2 k_1^-} x_2^- + i {\un k}_2 \cdot {\un x}_{12}} \, \mbox{Re} \,  \Bigg\langle \mbox{T} \, i \, (V_{\ul 1}^\dagger)^{ji} \notag \\ &  \times  \, \tr \Bigg[ \slashed{k_1} \, \bigg[  - \left( i \, \gamma^5 \, k_2^x - k_2^y \right) \,  \gamma^+ \, \gamma^-  + (i \, \gamma^5 \, k_1^x - k_1^y) \, \gamma^- \, \gamma^+ \bigg]  \,  \slashed{k_2} \,  \left( {\hat V}_{\un 2} \right)^{ij}  \Bigg] -  \mbox{c.c.}  \Bigg\rangle_{S_x = +1}  \, \Bigg|_{k_1^2 = k_2^2 =0, \, k_1^- = k_2^-}   . 
\end{align}
Integrating over $x_1^-, x_2^-$  yields
\begin{align}\label{Tev3}
& \delta T^{NS}_{10} (z s) =  \frac{g^2 \, (p^+)^2 }{8} \, \int d^2 x_2 \, \frac{d k_1^- \, d^2 k_1}{(2 \pi)^3 \, {\un k}_1^2 \, {\un k}_2^2} \, \frac{d^2 k_2}{(2 \pi)^2} \, e^{ i {\un k}_1 \cdot {\un x}_{21}} \, e^{ i {\un k}_2 \cdot {\un x}_{12}}  \\ &  \times \, \mbox{Re} \,  \Bigg\langle \mbox{T} \, i  \, (V_{\ul 1}^\dagger)^{ji} \, \tr \Bigg[ \slashed{k_1}   \, \bigg[ -  \left( i \, \gamma^5 \, k_2^x - k_2^y \right) \,  \gamma^+ \, \gamma^-  + (i \, \gamma^5 \, k_1^x - k_1^y) \, \gamma^- \, \gamma^+ \bigg]  \,  \slashed{k_2} \,  \left( {\hat V}_{\un 2} \right)^{ij}  \Bigg]  -  \mbox{c.c.} \Bigg\rangle_{S_x = +1}  \, \Bigg|_{k_1^2 = k_2^2 =0, \, k_1^- = k_2^-}  . \notag 
\end{align}

Next we evaluate the Dirac trace with the help of massless anti-BL spinors (since the mass does not contribute to the DLA evolution at hand),
\begin{align}\label{big_tr}
 & \tr \Bigg[ \slashed{k_1}   \, \bigg[  - \left( i \, \gamma^5 \, k_2^x - k_2^y \right) \,  \gamma^+ \, \gamma^-  + (i \, \gamma^5 \, k_1^x - k_1^y) \, \gamma^- \, \gamma^+ \bigg]  \,  \slashed{k_2} \,  \left( {\hat V}_{\un 2} \right)^{ij}  \Bigg]  \notag \\ & = \sum_{\chi_1, \chi_2} \, {\bar u}_{\chi_1} (k_1) \,  \bigg[  - \left( i \, \gamma^5 \, k_2^x - k_2^y \right) \,  \gamma^+ \, \gamma^-  + (i \, \gamma^5 \, k_1^x - k_1^y) \, \gamma^- \, \gamma^+ \bigg] u_{\chi_2} (k_2) \, {\bar u}_{\chi_2} (k_2) \, \left( {\hat V}_{\un 2} \right)^{ij} u_{\chi_1} (k_1) \notag \\ & = 16 \,  i \, k_1^- \, {\un k}_1 \cdot {\un k}_2 \,  \left( V_{\ul 2}^{pol, T} \right)^{ij},
\end{align}
where we have used 
\begin{align}
{\bar u}_{\chi_1} (k_2) \, \Big( \hat{V}_{\ul x} \Big) \, u_{\chi_2} (k_1) =
2 \sqrt{k_1^- k_2^{-}}\left( \delta_{\chi_1,  \chi_2} \, V_{\ul x} + \chi_1 \, \delta_{\chi_1,  \chi_2} \, V_{\ul x}^{pol, T} +  \ldots \right) 
\end{align}
and
\begin{align}
{\bar u}_{\chi_1} (k_1) \,  \bigg[ -  \left( i \, \gamma^5 \, k_2^x - k_2^y \right) \,  \gamma^+ \, \gamma^-  + (i \, \gamma^5 \, k_1^x - k_1^y) \, \gamma^- \, \gamma^+ \bigg] u_{\chi_2} (k_2) = 4 i \, \chi_1 \, \delta_{\chi_1,  \chi_2} \, {\un k}_1 \cdot {\un k}_2 + \ldots. 
\end{align}
In the last formula the ellipsis denote the energy-suppressed and off-diagonal in polarizations corrections, along with mass-dependent terms which do not contribute to the DLA evolution. 

Inserting \eq{big_tr} into \eq{Tev3} we get 
\begin{align}\label{Tev4}
\delta T^{NS}_{10} (z s) = - 2 g^2 \, (p^+)^2 \, \int d^2 x_2 \, \frac{d k_1^- \, k_1^- \,  d^2 k_1}{(2 \pi)^3 \, {\un k}_1^2 \, {\un k}_2^2} \, \frac{d^2 k_2}{(2 \pi)^2} \, e^{ i {\un k}_1 \cdot {\un x}_{21}} \, e^{ i {\un k}_2 \cdot {\un x}_{12}} \, {\un k}_1 \cdot {\un k}_2  \, \mbox{Re} \, \Bigg\langle \mbox{T} \, \tr \left[ V_{\ul 2}^{pol, T} \, V_{\ul 1}^\dagger \right] - \mbox{c.c.}  \Bigg\rangle_{S_x = +1} ,
\end{align}
or, equivalently, 
\begin{align}\label{Tev5}
\delta T^{NS}_{10} (z s) = \frac{\as \, N_c  }{2 \, \pi^2} \, \int \frac{d^2 x_2}{x_{21}^2} \, \frac{d z'}{z'} \, T^{NS}_{21} (z' s).
\end{align}

Inserting the proper DLA limits \cite{Itakura:2003jp,Kovchegov:2015pbl,Kovchegov:2016zex,Kovchegov:2017jxc} we arrive at the following evolution equation for the flavor non-singlet transversely polarized dipole amplitude:
\begin{align}\label{Tev6}
T^{NS}_{10} (z s) = T_{10}^{NS, \, (0)} (z s) + \frac{\as \, N_c  }{2 \, \pi} \, \int\limits_{\Lambda^2/s}^z \frac{d z'}{z'} \, 
\int\limits_{1/z' s}^{x_{10}^2 z/z'}  \frac{d x_{21}^2}{x_{21}^2} \  T^{NS}_{21} (z' s),
\end{align}
with the inhomogeneous term $T_{10}^{NS, \, (0)} (z s)$ to be found by performing a Born-level evaluation of the amplitude. Since our goal here is in establishing the small-$x$ asymptotics for the valence quark transversity, we will not need an explicit form of $T_{10}^{NS, \, (0)} (z s)$ since it does not affect the asymptotics.


\subsection{Solution of the evolution equations for the transversely polarized dipole}
\label{sec:sol}

Equation \eqref{Tev6} is mathematically identical to the evolution equation for the Reggeon amplitude derived in \cite{Itakura:2003jp} (see Eq.~(44) there). Defining new variables
\begin{align}
y = \ln ( z s x_{10}^2), \ \ \ \eta = \ln \frac{zs}{\Lambda^2}, \ \ \ y' = \ln ( z' s x_{21}^2), \ \ \ \eta' = \ln \frac{z's}{\Lambda^2},
\end{align}
we rewrite \eq{Tev6} as
\begin{align}\label{Tev7}
T^{NS} (\eta, y) = T^{NS, \, (0)} (\eta, y) + \frac{\as \, N_c  }{2 \, \pi} \, \int\limits_{0}^\eta d \eta' \, 
\int\limits_{0}^{y}  d y' \  T^{NS} (\eta', y').
\end{align}
The solution is obtained by performing a double Laplace-Mellin transform
\begin{align}\label{Mellin}
T^{NS} (\eta, y) = \int \frac{d \omega}{2 \pi i} \, \frac{d \lambda}{2 \pi i} \, e^{\omega \, \eta + \lambda \, y} \, T^{NS}_{\omega, \lambda} 
\end{align}
where the $\omega$ and $\lambda$ integrals run parallel to the imaginary axis to the right of all the singularities of the integrand. The Laplace-Mellin transform casts \eq{Tev7} in the following form:
\begin{align}\label{Tev8}
T^{NS}_{\omega, \lambda} = T_{\omega, \lambda}^{NS, \, (0)} + \frac{\as \, N_c  }{2 \, \pi} \, \frac{1}{\omega \, \lambda} \,   T^{NS}_{\omega, \lambda} .
\end{align}
Solving for $T^{NS}_{\omega, \lambda}$ we arrive at
\begin{align}
T^{NS}_{\omega, \lambda} = T_{\omega, \lambda}^{NS, \, (0)} \, \frac{1}{1 - \frac{\as \, N_c  }{2 \, \pi} \, \frac{1}{\omega \, \lambda}}
\end{align}
such that 
\begin{align}\label{solution1}
T^{NS} (\eta, y) = \int \frac{d \omega}{2 \pi i} \, \frac{d \lambda}{2 \pi i} \, e^{\omega \, \eta + \lambda \, y} \, T_{\omega, \lambda}^{NS, \, (0)} \, \frac{1}{1 - \frac{\as \, N_c  }{2 \, \pi} \, \frac{1}{\omega \, \lambda}}. 
\end{align}
Next we integrate over $\omega$ assuming that the double Laplace-Mellin transform of the inhomogeneous term does not generate any poles to the right of the pole in the denominator of \eq{solution1}. This gives the leading asymptotics
\begin{align}\label{solution2}
T^{NS} (\eta, y) = \int \frac{d \lambda}{2 \pi i}  \, \frac{\as \, N_c  }{2 \, \pi} \, \frac{1}{\lambda} \, T_{\frac{\as \, N_c  }{2 \, \pi} \, \frac{1}{\lambda}, \lambda}^{NS, \, (0)} \, \exp \left\{ \frac{\as \, N_c  }{2 \, \pi} \, \frac{1}{\lambda} \, \eta + \lambda \, y \right\} .
\end{align}
Again assuming that the singularities of the inhomogeneous term are not important, we distort the $\lambda$-contour into its steepest descent form, going through the saddle point at 
\begin{align}
\lambda_{s.p.} = \sqrt{\frac{\as \, N_c  }{2 \, \pi} \, \frac{\eta}{y}}.
\end{align}
The integral is then dominated by $\lambda = \lambda_{s.p.}$ such that
\begin{align}
T^{NS} (\eta, y) \sim \exp \left\{ 2\, \sqrt{\frac{\as \, N_c  }{2 \, \pi} } \, \sqrt{\eta \, y} \right\}.
\end{align}
At high energies $\eta \approx y \sim \ln (z s)$, such that the asymptotics of the flavor non-singlet transversely polarized dipole amplitude is
\begin{align}\label{Tasym}
T^{NS}_{10} (zs) \sim (z s)^{ 2\, \sqrt{\frac{\as \, N_c  }{2 \, \pi} }}.
\end{align}

We realize that we are indeed in the situation described by \eq{sample_asym} with $C = 2 \, \sqrt{N_c/(2 \pi)} >0$. This means that \eq{trTMD8NS} does give us the leading small-$x$ asymptotics for valence transversity. Employing \eq{Tasym} in \eq{trTMD8NS} we conclude that 
\begin{align}\label{trans_asym}
h_{1T}^{NS} (x, k_T^2) \sim h_{1T}^{\perp \, NS} (x, k_T^2) \sim \left( \frac{1}{x} \right)^{-1 + 2 \, \sqrt{\frac{\as \, N_c}{2 \, \pi}} } .
\end{align}
This is our main result, giving us the asymptotics of valence quark transversity. As mentioned in the Introduction, it agrees with one of the small-$x$ asymptotics for transversity found in \cite{Kirschner:1996jj}.

The power of $1/x$ in \eq{trans_asym} is much smaller than the intercept of the leading-order Balitsky--Fadin--Kuraev--Lipatov (BFKL) \cite{Kuraev:1977fs,Balitsky:1978ic} evolution equation, whole solution determines the small-$x$ asymptotics for unpolarized TMD. This power is also much smaller than the quark helicity intercept found in \cite{Kovchegov:2016weo,Kovchegov:2017jxc}. Hence the Soffer bound \cite{Soffer:1994ww} appears to be easily satisfied by \eq{trans_asym} at small $x$. 

For $\as = 0.3$ we get from \eq{trans_asym} with $N_c =3$
\begin{align}\label{trTMDnumber}
h_{1T}^{NS} (x, k_T^2) \sim h_{1T}^{\perp \, NS} (x, k_T^2) \sim x^{0.243}.
\end{align}


\subsection{An alternative derivation}

One may argue that the above operator method of obtaining the small-$x$ asymptotics of the valence quark transversity, while powerful, lacks physics insight. To generate the latter, we will now re-derive the result \eqref{trans_asym} using the more conventional ladder diagram technique. 

\begin{figure}[ht]
\begin{center}
\includegraphics[width= 0.7 \textwidth]{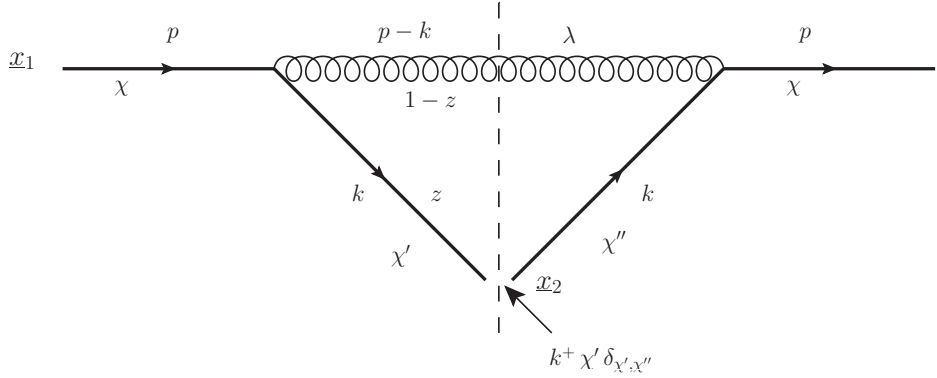} 
\caption{Diagrammatic representation of transversity TMD at the lowest order. The dashed line separates the light-cone wave function from its complex conjugate.}
\label{fig:wf}
\end{center}
\end{figure}

Returning to the conventional BL spinors \cite{Lepage:1980fj,Brodsky:1997de} and using the fact that 
\begin{align}\label{5+1}
{\bar u}_\chi (p) \, \thalf \, \gamma^5 \, \gamma^+ \, \gamma^1 \, u_{\chi'} (k) = \sqrt{p^+ \, k^+} \, \chi \, \delta_{\chi, \chi'}
\end{align}
for these spinors, we can calculate the leading-order transversity TMD or, equivalently, one step of transversity evolution using the diagram squared pictured in \fig{fig:wf}. Here we assume that the incoming quark in \fig{fig:wf} originates somewhere in the transversely polarized proton, and is, in turn, transversely polarized. The soft outgoing quark is the parton whose distribution we are interested in.

In momentum space we need (with the factor of $k^+$ coming from the use of \eq{5+1})
\begin{align}
k^+ \, \sum_{\lambda, \chi'} \, \chi' \, |\psi_{\lambda, \chi, \chi'}|^2,
\end{align}
where $\psi_{\lambda, \chi, \chi'}$ is the light-cone wave function, the square of which is pictured in \fig{fig:wf}. Working in a frame with ${\un p} =0$ we have (see Eq.~(9) in \cite{Kovchegov:2012ga})
\begin{align}
 \psi_{\lambda \chi \chi'}^a ( {\un k} , z) = \frac{g \, t^a}{{\un k}^2 + {\tilde m}^2} 
 \bigg[ {\un \epsilon}_{\lambda}^* \cdot {\un k}  \, \bigg(
 (1+z) \, \delta_{\chi, \chi'} + \lambda \, (1-z) \, \delta_{\chi, -\chi'} \bigg) 
 - \frac{\tilde m}{\sqrt 2} \, (1-z) \, \chi \,
 \big (\delta_{\chi , \chi'} - \lambda \, \delta_{\chi, -\chi'} \big) \bigg] \;, 
 \label{eq-momentumwavefn}
\end{align}
where
\begin{equation}
 \nonumber
 \tilde m \equiv (1-z) m
\end{equation}
and $z = k^+/p^+$. After some algebra one readily obtains
\begin{align}
k^+ \, \sum_{\lambda, \chi'} \, \chi' \, |\psi_{\lambda, \chi, \chi'} ( {\un k} , z)|^2   = \frac{4 g^2 \, C_F \, z \, k_\perp^2}{\left[ {k}_\perp^2 + (1-z)^2 \, {m}^2 \right]^2} \, k^+ \, \chi . 
\end{align}
It has to be multiplied by the phase-space factor \cite{Kovchegov:2012mbw}
\begin{align}
\frac{d k^+ \, d^2 k_\perp}{2 k^+ \, (2 \pi)^3} \, \frac{p^+}{p^+ - k^+}
\end{align}
which gives
\begin{align}\label{transverse1}
\frac{2 g^2 \, C_F \, z \, k_\perp^2}{\left[ {k}_\perp^2 + (1-z)^2 \, {m}^2 \right]^2} \, \chi \, \frac{d z \, d^2 k_\perp}{(1-z) \,  (2 \pi)^3}.
\end{align}

Comparing this result with the Eq.~(50) in \cite{Meissner:2007rx} we read off
\begin{align}
h_{1T}^q = \frac{2 g^2 \, C_F \, z \, k_\perp^2}{\left[ {k}_\perp^2 + (1-z)^2 \, {m}^2 \right]^2} \, \frac{1}{(1-z) \,  (2 \pi)^3}, \ \ \ \ \ h_{1T}^{\perp q} = 0,
\end{align}
in agreement with Eqs.~(B9) and (B10) in the same reference. 

At small $z$ and for $k_\perp \gg m$ \eq{transverse1} becomes (dropping the factor of $\chi$)
\begin{align}
\int\limits_{\Lambda^2/s}^1 dz \, z \, \int\limits_{m}^{\sqrt{s}} \frac{d^2 k_\perp}{k_\perp^2} \, \frac{2 g^2 \, C_F}{(1-z) \,  (2 \pi)^3}.
\end{align}
Note that the $z$-integral will become logarithmic once we multiply this by the factor $\sim 1/(z^2 \, s^2)$: if the splitting of \fig{fig:wf} happened in the projectile dipole, then the factor of $\sim 1/(z^2 \, s^2)$ would have been responsible for the transverse spin-dependent interaction of the soft quark from \fig{fig:wf} with the transversely polarized proton target, which is doubly-suppressed at high energy. 

\begin{figure}[ht]
\begin{center}
\includegraphics[width= 0.4 \textwidth]{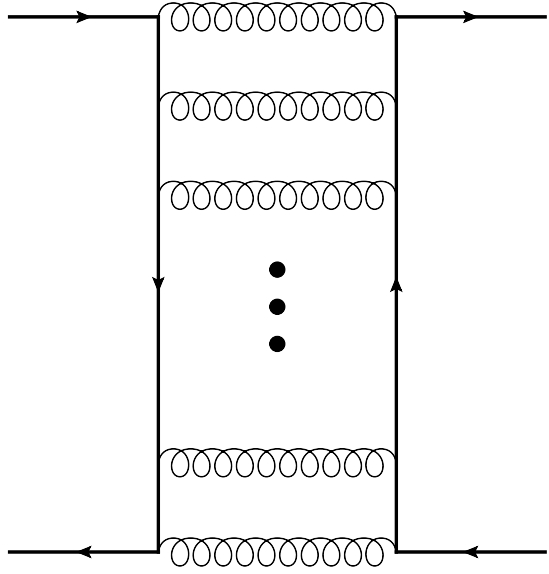} 
\caption{Double-logarithmic evolution of the transversity TMD at large $N_c$, the lowest order for which is given by \fig{fig:wf}. }
\label{fig:quark_ladder}
\end{center}
\end{figure}

The same calculation can be repeated in the transverse coordinate space. Starting with the coordinate-space wave function 
\begin{align}
 \psi_{\lambda \chi \chi'}^a ({\un x}_{21}, z) = & 
\frac{g \, t^a}{2\pi} \, {\tilde m}  \, \bigg\{ 
  i \, {\un \epsilon}_\lambda^* \cdot \frac{{\un x}_{21}}{x_{21}} \, 
 K_1 ( \tilde m \, x_{21}) \bigg[ (1+z) \, \delta_{\chi,\chi'} +
 \lambda \, (1-z) \, \delta_{\chi, -\chi'} \bigg]  \notag \\ & - \frac{\chi \, (1-z)}{\sqrt 2} \, K_0 (\tilde m \, x_{21}) \, \bigg[ \delta_{\chi, \chi'} - \lambda \,
 \delta_{\chi, - \chi'} \bigg] \bigg\} \label{eq-coord wavefn}
\end{align} 
and assuming the small-$z$ limit from the start we arrive at
\begin{align}\label{transverse2}
k^+ \, \sum_{\lambda, \chi'} \, \chi' \, |\psi_{\lambda, \chi, \chi'}|^2 \, \frac{dz \, d^2 x_{21}}{4 \pi} = \chi \, \frac{\as \, C_F}{\pi} \, m^2 \, K_1 ( m \, x_{21})^2 \, z \, dz \, d x_{21}^2 . 
\end{align}
Somewhat surprisingly, in arriving at \eq{transverse2} we had to drop out the terms linear in ${\un x}_{21}$ as integrating out to zero after the angular part of the $d^2 x_{21}$ integration. One could also argue that those dropped terms were $\chi$-independent.

In the DLA region of $z \ll 1$ and $x_{21} \ll 1/m$ we obtain the following evolution kernel
\begin{align}\label{tr_kernel}
K^{q {\bar q} \to q {\bar q}} = \frac{\as \, C_F}{\pi} \, \int\limits_{\Lambda^2/s}^1 \frac{dz}{z} \, \int \frac{d x_{21}^2}{x_{21}^2}, 
\end{align}
if we multiply \eq{transverse2} by $1/z^2$ coming from the transverse spin-dependent interaction with the projectile. At large-$N_c$ this is exactly the kernel of \eq{Tev6}. In fact, \eq{Tev6} can be constructed by successive iterations of the soft-quark emission of \fig{fig:wf}. This results in the ladder with quarks in the $t$-channel pictured in \fig{fig:quark_ladder}. In general, the DLA evolution for transversity is not limited to ladder diagrams \cite{Kirschner:1996jj}: however, at large $N_c$ only ladder diagrams remain in the evolution. We, therefore, conclude that our \eq{Tev6} corresponds to quark ladder of \fig{fig:quark_ladder} (cf. the Reggeon evolution of \cite{Itakura:2003jp}).


\section{Conclusions}

In this work we have applied the operator formalism developed in \cite{Kovchegov:2018znm} to establishing the small-$x$ asymptotics of the valence quark transversity at small $x$. The result for the asymptotics is given in \eq{trans_asym} above. Our calculation demonstrates that indeed the formalism of \cite{Kovchegov:2018znm}, perhaps with minor modifications for the sub-sub-eikonal case, can be applied to other quark TMDs to determine their small-$x$ asymptotics. 

One a more physical side, the power of $1/x$ we obtain in \eq{trans_asym} is rather low. From using \eq{trans_asym} in \eq{tensor} it appears likely that one would find only a rather modest amount of the proton tensor charge residing at small-$x$. Hence it is conceivable that the small-$x$ region would not help resolve the ``transverse spin puzzle" outlined in \cite{Radici:2018iag}. However, a detailed phenomenological analysis is needed for a definitive conclusion.


\section*{Acknowledgments}

We acknowledge useful communications with Daniel Boer, Aurore Courtoy, Gary Goldstein, Daniel Pitonyak, Alexey Prokudin, Marco Radici, and Jian Zhou. 

This material is based upon work supported by the U.S. Department of
Energy, Office of Science, Office of Nuclear Physics under Award
Number DE-SC0004286 (YK) and DOE Contract No. DE-AC52-06NA25396 (MS).  MS
received additional support from the U.S. Department of Energy, Office
of Science under the DOE Early Career Program.\\




\appendix
\section{Other possible contributions to transversely polarized quark Wilson line}
\label{A}

\begin{figure}[ht]
\begin{center}
\includegraphics[width=  \textwidth]{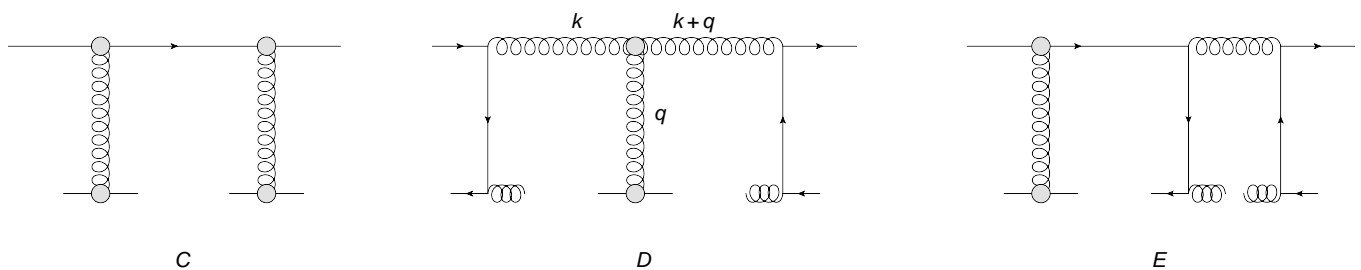} 
\caption{Additional diagrams potentially contributing to the transversely polarized fundamental Wilson line in the
  quasi-classical approximation in $A^-=0$ gauge. The gray circles 
  denote the spin-dependent sub-eikonal scattering.}
\label{more}
\end{center}
\end{figure}

Since the transverse spin dependence is sub-sub-eikonal, one has to consider other possible sub-sub-eikonal contributions, in addition to those considered in Sec.~\ref{sec:trWil}. In addition to the diagrams in \fig{vpolT} one has to include the diagrams shown in \fig{more}, where the $t$-channel gluon interactions are taken at the sub-eikonal level each, such that, e.g., two $t$-channel gluon exchanges in the diagram C combine to give a sub-sub-eikonal interaction. 

Luckily all of the diagrams in \fig{more} are independent of the transverse spin. Using the transverse matrix elements for BL spinors  \cite{Lepage:1980fj,Brodsky:1997de},
\begin{subequations}
\begin{align}
& {\bar u}_\chi (p-k) \, \gamma^+ \, u_{\chi'} (p) = 2 \, \delta_{\chi, \chi'} \, \sqrt{p^+ \, (p^+ - k^+)}  , \\
& {\bar u}_\chi (p-k) \, \gamma^- \, u_{\chi'} (p) = \frac{2}{\sqrt{p^+ \, (p^+ - k^+)}} \, \left[ \delta_{\chi, \chi'} \, \left( p_x \, (p-k)_x + (m + i p_y \, \chi) \, (m - i (p-k)_y \, \chi) )\right) \right. \notag \\ & \hspace*{5cm} \left. + \delta_{\chi, - \chi'} \, \left( m p_x \chi + i p_y (p-k)_x - m (p-k)_x \chi - i p_x (p-k)_y \right) \right] , \\
& {\bar u}_\chi (p-k) \, \gamma^i \, u_{\chi'} (p) = \sqrt{p^+ \, (p^+ - k^+)} \, \left[ \delta_{\chi, \chi'} \, \left( \frac{p^i}{p^+} + \frac{(p-k)^i}{p^+ - k^+} - i m \left( \frac{1}{p^+} - \frac{1}{p^+ - k^+} \right) \, \chi \, \delta^{i2}   \right) \right. \notag \\ & \hspace*{5cm} \left.  + \delta_{\chi, - \chi'} \,  \left( i \, \epsilon^{ij} \, \left( \frac{p^j}{p^+} - \frac{(p-k)^j}{p^+ - k^+} \right) - m \left( \frac{1}{p^+} - \frac{1}{p^+ - k^+} \right) \, \chi \, \delta^{i1} \right) \right] , 
\end{align}
\end{subequations}
one can show that the sub-eikonal gluon field only contains $\delta_{\chi, \chi'}$ and $\delta_{\chi, - \chi'}$, in contrast to the sub-sub-eikonal gluon field in Eqs.~\eqref{fields}. Here $\chi$ and $\chi'$ are transverse polarizations of the quark in the plus-moving target proton before and after the emission of the gluon field.  (The eikonal gluon field is proportional to $\delta_{\chi, \chi'}$, as usual.) There are no multiplicative factors of $\chi$ and $\chi'$ in the eikonal and sub-eikonal gluon fields, only $\delta_{\chi, \chi'}$ and $\delta_{\chi, - \chi'}$. This means that one cannot get $\chi \, \delta_{\chi, \chi'}$ and the transverse spin dependence out of diagram C from \fig{more}. The same is true for diagram E: the two-quark exchange at sub-eikonal level only contains $\delta_{\chi, \chi'}$ and $\delta_{\chi, - \chi'}$ (see the first line of \eq{Vpolq2} above). Combining this with the sub-eikonal contribution to the $t$-channel gluon exchange, also containing $\delta_{\chi, \chi'}$ and $\delta_{\chi, - \chi'}$ only, without any factors of $\chi$ or $\chi'$, one again cannot generate $\chi \, \delta_{\chi, \chi'}$ and the transverse spin dependence. 

The sub-eikonal ($A^- =0$ gauge) gluon field, multiplied by the triple-gluon vertex, contributes the following to the diagram D:
\begin{align}\label{tr_field}
\sim \delta_{\lambda, \lambda'} \, \left[ i \, \lambda \, {\un q} \times {\un A} (q) + (2 {\un k} + {\un q}) \cdot {\un A} (q) + (2 k^- + q^-) \, A^+ (q) \right] .
\end{align}
As far as the $\lambda$-dependence is concerned, the only difference between the eikonal vertex ($\sim  \delta_{\lambda, \lambda'}$) and the expression in \eq{tr_field} is in the factor of $\lambda$ in the square brackets of the latter. Even with this correction, the contribution of the diagram D in \fig{more} is proportional to a linear combination of $\delta_{\chi, \chi'}$ and $\delta_{\chi, - \chi'}$, and, hence, does not depend on the transverse spin. 

We conclude that the diagrams in \fig{more} do not contribute transverse spin dependence at the sub-sub-eikonal level in question, and are not needed in the analysis of Sec.~\ref{sec:trWil}.



\begin{thebibliography}{10}

\bibitem{Kirschner:1996jj}
R.~Kirschner, L.~Mankiewicz, A.~Schafer and L.~Szymanowski, \emph{{Small x
  behavior of the chirally odd parton distribution $h_1 (x, Q^2)$}},
  \href{https://doi.org/10.1007/s002880050412}{\emph{Z. Phys.} {\bfseries C74}
  (1997) 501--508}, [\href{https://arxiv.org/abs/hep-ph/9606267}{{\ttfamily
  hep-ph/9606267}}].

\bibitem{Ralston:1979ys}
J.~P. Ralston and D.~E. Soper, \emph{{Production of Dimuons from High-Energy
  Polarized Proton Proton Collisions}},
  \href{https://doi.org/10.1016/0550-3213(79)90082-8}{\emph{Nucl. Phys.}
  {\bfseries B152} (1979) 109}.

\bibitem{Cortes:1991ja}
J.~Cortes, B.~Pire and J.~Ralston, \emph{{Measuring the transverse polarization
  of quarks in the proton}},
  \href{https://doi.org/10.1007/BF01565099}{\emph{Z.Phys.} {\bfseries C55}
  (1992) 409--416}.

\bibitem{Jaffe:1991ra}
R.~L. Jaffe and X.-D. Ji, \emph{{Chiral odd parton distributions and Drell-Yan
  processes}}, \href{https://doi.org/10.1016/0550-3213(92)90110-W}{\emph{Nucl.
  Phys.} {\bfseries B375} (1992) 527--560}.

\bibitem{Barone:2001sp}
V.~Barone, A.~Drago and P.~G. Ratcliffe, \emph{{Transverse polarisation of
  quarks in hadrons}},
  \href{https://doi.org/10.1016/S0370-1573(01)00051-5}{\emph{Phys. Rept.}
  {\bfseries 359} (2002) 1--168},
  [\href{https://arxiv.org/abs/hep-ph/0104283}{{\ttfamily hep-ph/0104283}}].

\bibitem{Anselmino:2007fs}
M.~Anselmino, M.~Boglione, U.~D'Alesio, A.~Kotzinian, F.~Murgia et~al.,
  \emph{{Transversity and Collins functions from SIDIS and e+ e- data}},
  \href{https://doi.org/10.1103/PhysRevD.75.054032}{\emph{Phys.Rev.} {\bfseries
  D75} (2007) 054032}, [\href{https://arxiv.org/abs/hep-ph/0701006}{{\ttfamily
  hep-ph/0701006}}].

\bibitem{Radici:2018iag}
M.~Radici and A.~Bacchetta, \emph{{First Extraction of Transversity from a
  Global Analysis of Electron-Proton and Proton-Proton Data}},
  \href{https://doi.org/10.1103/PhysRevLett.120.192001}{\emph{Phys. Rev. Lett.}
  {\bfseries 120} (2018) 192001},
  [\href{https://arxiv.org/abs/1802.05212}{{\ttfamily 1802.05212}}].

\bibitem{Bacchetta:2012ty}
A.~Bacchetta, A.~Courtoy and M.~Radici, \emph{{First extraction of valence
  transversities in a collinear framework}},
  \href{https://doi.org/10.1007/JHEP03(2013)119}{\emph{JHEP} {\bfseries 03}
  (2013) 119}, [\href{https://arxiv.org/abs/1212.3568}{{\ttfamily 1212.3568}}].

\bibitem{Kovchegov:2018znm}
Y.~V. Kovchegov and M.~D. Sievert, \emph{{Small-$x$ Helicity Evolution: an
  Operator Treatment}},  \href{https://arxiv.org/abs/1808.09010}{{\ttfamily
  1808.09010}}.

\bibitem{Kovchegov:2015pbl}
Y.~V. Kovchegov, D.~Pitonyak and M.~D. Sievert, \emph{{Helicity Evolution at
  Small-x}}, \href{https://doi.org/10.1007/JHEP01(2016)072}{\emph{JHEP}
  {\bfseries 01} (2016) 072},
  [\href{https://arxiv.org/abs/1511.06737}{{\ttfamily 1511.06737}}].

\bibitem{Kovchegov:2016weo}
Y.~V. Kovchegov, D.~Pitonyak and M.~D. Sievert, \emph{{Small-$x$ asymptotics of
  the quark helicity distribution}},
  \href{https://doi.org/10.1103/PhysRevLett.118.052001}{\emph{Phys. Rev. Lett.}
  {\bfseries 118} (2017) 052001},
  [\href{https://arxiv.org/abs/1610.06188}{{\ttfamily 1610.06188}}].

\bibitem{Kovchegov:2016zex}
Y.~V. Kovchegov, D.~Pitonyak and M.~D. Sievert, \emph{{Helicity Evolution at
  Small $x$: Flavor Singlet and Non-Singlet Observables}},
  \href{https://doi.org/10.1103/PhysRevD.95.014033}{\emph{Phys. Rev.}
  {\bfseries D95} (2017) 014033},
  [\href{https://arxiv.org/abs/1610.06197}{{\ttfamily 1610.06197}}].

\bibitem{Kovchegov:2017jxc}
Y.~V. Kovchegov, D.~Pitonyak and M.~D. Sievert, \emph{{Small-$x$ Asymptotics of
  the Quark Helicity Distribution: Analytic Results}},
  \href{https://arxiv.org/abs/1703.05809}{{\ttfamily 1703.05809}}.

\bibitem{Kovchegov:2017lsr}
Y.~V. Kovchegov, D.~Pitonyak and M.~D. Sievert, \emph{{Small-$x$ Asymptotics of
  the Gluon Helicity Distribution}},
  \href{https://doi.org/10.1007/JHEP10(2017)198}{\emph{JHEP} {\bfseries 10}
  (2017) 198}, [\href{https://arxiv.org/abs/1706.04236}{{\ttfamily
  1706.04236}}].

\bibitem{Kirschner:1983di}
R.~Kirschner and L.~Lipatov, \emph{{Double Logarithmic Asymptotics and Regge
  Singularities of Quark Amplitudes with Flavor Exchange}},
  \href{https://doi.org/10.1016/0550-3213(83)90178-5}{\emph{Nucl.Phys.}
  {\bfseries B213} (1983) 122--148}.

\bibitem{Kirschner:1985cb}
R.~Kirschner, \emph{{Regge Asymptotics of Scattering Amplitudes in the
  Logarithmic Approximation of {QCD}}},
  \href{https://doi.org/10.1007/BF01559604}{\emph{Z. Phys.} {\bfseries C31}
  (1986) 135}.

\bibitem{Kirschner:1994vc}
R.~Kirschner, \emph{{Regge asymptotics of scattering with flavor exchange in
  QCD}}, \href{https://doi.org/10.1007/BF01624588}{\emph{Z.Phys.} {\bfseries
  C67} (1995) 459--466},
  [\href{https://arxiv.org/abs/hep-th/9404158}{{\ttfamily hep-th/9404158}}].

\bibitem{Kirschner:1994rq}
R.~Kirschner, \emph{{Reggeon interactions in perturbative QCD}},
  \href{https://doi.org/10.1007/BF01556138}{\emph{Z.Phys.} {\bfseries C65}
  (1995) 505--510}, [\href{https://arxiv.org/abs/hep-th/9407085}{{\ttfamily
  hep-th/9407085}}].

\bibitem{Griffiths:1999dj}
S.~Griffiths and D.~Ross, \emph{{Studying the perturbative Reggeon}},
  \href{https://doi.org/10.1007/s100529900240}{\emph{Eur.Phys.J.} {\bfseries
  C12} (2000) 277--286},
  [\href{https://arxiv.org/abs/hep-ph/9906550}{{\ttfamily hep-ph/9906550}}].

\bibitem{Itakura:2003jp}
K.~Itakura, Y.~V. Kovchegov, L.~McLerran and D.~Teaney, \emph{{Baryon stopping
  and valence quark distribution at small x}},
  \href{https://doi.org/10.1016/j.nuclphysa.2003.10.016}{\emph{Nucl. Phys.}
  {\bfseries A730} (2004) 160--190},
  [\href{https://arxiv.org/abs/hep-ph/0305332}{{\ttfamily hep-ph/0305332}}].

\bibitem{Bartels:2003dj}
J.~Bartels and M.~Lublinsky, \emph{{Quark anti-quark exchange in gamma* gamma*
  scattering}},
  \href{https://doi.org/10.1088/1126-6708/2003/09/076}{\emph{JHEP} {\bfseries
  0309} (2003) 076}, [\href{https://arxiv.org/abs/hep-ph/0308181}{{\ttfamily
  hep-ph/0308181}}].

\bibitem{Chirilli:2013kca}
G.~A. Chirilli and Y.~V. Kovchegov, \emph{{Solution of the NLO BFKL Equation
  and a Strategy for Solving the All-Order BFKL Equation}},
  \href{https://doi.org/10.1007/JHEP06(2013)055}{\emph{JHEP} {\bfseries 1306}
  (2013) 055}, [\href{https://arxiv.org/abs/1305.1924}{{\ttfamily 1305.1924}}].

\bibitem{Lepage:1980fj}
G.~P. Lepage and S.~J. Brodsky, \emph{Exclusive processes in perturbative
  quantum chromodynamics}, {\emph{Phys. Rev.} {\bfseries D22} (1980) 2157}.

\bibitem{Kovchegov:2012ga}
Y.~V. Kovchegov and M.~D. Sievert, \emph{{A New Mechanism for Generating a
  Single Transverse Spin Asymmetry}},
  \href{https://doi.org/10.1103/PhysRevD.86.034028}{\emph{Phys.Rev.} {\bfseries
  D86} (2012) 034028}, [\href{https://arxiv.org/abs/1201.5890}{{\ttfamily
  1201.5890}}].

\bibitem{Meissner:2007rx}
S.~Meissner, A.~Metz and K.~Goeke, \emph{{Relations between generalized and
  transverse momentum dependent parton distributions}},
  \href{https://doi.org/10.1103/PhysRevD.76.034002}{\emph{Phys. Rev.}
  {\bfseries D76} (2007) 034002},
  [\href{https://arxiv.org/abs/hep-ph/0703176}{{\ttfamily hep-ph/0703176}}].

\bibitem{Kovchegov:2015zha}
Y.~V. Kovchegov and M.~D. Sievert, \emph{{Calculating TMDs of a Large Nucleus:
  Quasi-Classical Approximation and Quantum Evolution}},
  \href{https://doi.org/10.1016/j.nuclphysb.2015.12.008}{\emph{Nucl. Phys.}
  {\bfseries B903} (2016) 164--203},
  [\href{https://arxiv.org/abs/1505.01176}{{\ttfamily 1505.01176}}].

\bibitem{Gribov:1984tu}
L.~V. Gribov, E.~M. Levin and M.~G. Ryskin, \emph{{Semihard Processes in QCD}},
  {\emph{Phys. Rept.} {\bfseries 100} (1983) 1--150}.

\bibitem{Iancu:2003xm}
E.~Iancu and R.~Venugopalan, \emph{The color glass condensate and high energy
  scattering in {QCD}},  \href{https://arxiv.org/abs/hep-ph/0303204}{{\ttfamily
  hep-ph/0303204}}.

\bibitem{Weigert:2005us}
H.~Weigert, \emph{Evolution at small {$x_{bj}$: The Color Glass Condensate}},
  {\emph{Prog. Part. Nucl. Phys.} {\bfseries 55} (2005) 461--565},
  [\href{https://arxiv.org/abs/hep-ph/0501087}{{\ttfamily hep-ph/0501087}}].

\bibitem{Jalilian-Marian:2005jf}
J.~Jalilian-Marian and Y.~V. Kovchegov, \emph{Saturation physics and deuteron
  gold collisions at {RHIC}}, {\emph{Prog. Part. Nucl. Phys.} {\bfseries 56}
  (2006) 104--231}, [\href{https://arxiv.org/abs/hep-ph/0505052}{{\ttfamily
  hep-ph/0505052}}].

\bibitem{Gelis:2010nm}
F.~Gelis, E.~Iancu, J.~Jalilian-Marian and R.~Venugopalan, \emph{{The Color
  Glass Condensate}},
  \href{https://doi.org/10.1146/annurev.nucl.010909.083629}{\emph{Ann.Rev.Nucl.Part.Sci.}
  {\bfseries 60} (2010) 463--489},
  [\href{https://arxiv.org/abs/1002.0333}{{\ttfamily 1002.0333}}].

\bibitem{Albacete:2014fwa}
J.~L. Albacete and C.~Marquet, \emph{{Gluon saturation and initial conditions
  for relativistic heavy ion collisions}},
  \href{https://doi.org/10.1016/j.ppnp.2014.01.004}{\emph{Prog.Part.Nucl.Phys.}
  {\bfseries 76} (2014) 1--42},
  [\href{https://arxiv.org/abs/1401.4866}{{\ttfamily 1401.4866}}].

\bibitem{Kovchegov:2012mbw}
Y.~V. Kovchegov and E.~Levin, \emph{{Quantum chromodynamics at high energy}},
  vol.~33.
\newblock Cambridge University Press, 2012.

\bibitem{Mueller:2012bn}
A.~Mueller and S.~Munier, \emph{{$p_{\perp}$-broadening and production
  processes versus dipole/quadrupole amplitudes at next-to-leading order}},
  \href{https://doi.org/10.1016/j.nuclphysa.2012.08.005}{\emph{Nucl.Phys.}
  {\bfseries A893} (2012) 43--86},
  [\href{https://arxiv.org/abs/1206.1333}{{\ttfamily 1206.1333}}].

\bibitem{Brodsky:1997de}
S.~J. Brodsky, H.-C. Pauli and S.~S. Pinsky, \emph{{Quantum chromodynamics and
  other field theories on the light cone}},
  \href{https://doi.org/10.1016/S0370-1573(97)00089-6}{\emph{Phys.Rept.}
  {\bfseries 301} (1998) 299--486},
  [\href{https://arxiv.org/abs/hep-ph/9705477}{{\ttfamily hep-ph/9705477}}].

\bibitem{Kuraev:1977fs}
E.~A. Kuraev, L.~N. Lipatov and V.~S. Fadin, \emph{{The Pomeranchuk
  singlularity in non-Abelian gauge theories}}, {\emph{Sov. Phys. JETP}
  {\bfseries 45} (1977) 199--204}.

\bibitem{Balitsky:1978ic}
I.~Balitsky and L.~Lipatov, \emph{{The Pomeranchuk Singularity in Quantum
  Chromodynamics}}, {\emph{Sov.J.Nucl.Phys.} {\bfseries 28} (1978) 822--829}.

\bibitem{Soffer:1994ww}
J.~Soffer, \emph{{Positivity constraints for spin dependent parton
  distributions}},
  \href{https://doi.org/10.1103/PhysRevLett.74.1292}{\emph{Phys. Rev. Lett.}
  {\bfseries 74} (1995) 1292--1294},
  [\href{https://arxiv.org/abs/hep-ph/9409254}{{\ttfamily hep-ph/9409254}}].

\end{thebibliography}

\providecommand{\href}[2]{#2}\begingroup\raggedright\endgroup

\end{document}